\documentclass[12pt]{article}
\usepackage{hhline}
\usepackage{latexsym,graphicx}
\usepackage{mathrsfs}
\usepackage{subfigure}
\usepackage{amssymb}
\usepackage{amsmath}
\usepackage{amscd}
\usepackage{amsthm}
\usepackage{float}
\usepackage[left=2cm,top=2.5cm,right=2.5cm,bottom=1.5cm]{geometry}   
\usepackage{xcolor}
 
\begin{document}
\begin{center}
\large{\bf{Some cosmological features of 4D Gauss-Bonnet gravity with varying cosmological constant }} \\
\vspace{10mm}
\normalsize{Nasr Ahmed$^{1,2}$, Ajab A. Alfreedi$^1$ \& Alaa A. Alzulaibani$^1$ }\\
\vspace{5mm}
\small{\footnotesize $^1$ Mathematics and Statistics Department, Faculty of Science, Taibah University, KSA.} \\
\small{\footnotesize $^2$ Astronomy Department, National Research Institute of Astronomy and Geophysics, Helwan, Cairo, Egypt\footnote{abualansar@gmail.com}} \\
\end{center}  
\date{}
\begin{abstract}
We explore some cosmological features of the newly suggested 4D Gauss-Bonnet gravity through two different models assuming a varying cosmological constant $\Lambda(t)$. Observational constraints, such as the cosmic transit and the flat curvature, have been considered in constructing the models. The cosmology in the current work has been probed using a given scale factor derived from the desired cosmic behavior which is the inverse of the usual viewpoint. Several models for $\Lambda(t)$ have been proposed in the literature, we use two different ansatze of varying $\Lambda$ and compare the obtained result. We have found that the second ansatz for the varying $\Lambda$, expressed  in terms of the Hubble parameter $H$, gives better results than the first one. The sound speed causality condition along with all nonlinear energy conditions are satisfied for the case of $\Lambda(H)$. The cosmographic parameters have been investigated.

\end{abstract}

\section{Introduction and motivation}

The late-time cosmic acceleration has been a major challenging problem in cosmology and theoretical physics since its discovery in 1998 \cite{1}. One popular idea to explain this acceleration within the framework of general relativity is the unknown dark energy which can be simply represented by the cosmological constant. However, in order for the cosmological constant to play this role, its value must be incredibly small. The value of the cosmological constant expected from particle physics, through vacuum energy, is approximately 50 orders of magnitude larger than the observed value extracted from astrophysics assuming general relativity. A possible explanation is that an alternative modified gravity theory is required on cosmological scale with no need to assume dark energy \cite{2,3}.\par
Several modified gravity theories have been developed as an alternative to dark energy \cite{6}-\cite{12}, a good review has been given in \cite{4}. Examples include $f(R)$ gravity \cite{39} where $R$ is the Ricci scalar, Gauss-Bonnet gravity \cite{noj8} and f(T) gravity \cite{torsion} where $T$ is the torsion scalar. In Gauss-Bonnet gravity, the Ricci scalar $R$ in the action has been replaced by the Gauss-Bonnet term $G=R^2-4R^{\mu\nu}R_{\mu\nu}+R^{\mu\nu\rho\delta}R_{\mu\nu\rho\delta}$. $f(R,T)$ gravity \cite{5,5a}, is a generalization of $f(R)$ gravity where $T$ is the trace of the energy momentum tensor. Dynamical scalar fields is another explanation approach \cite{13}-\cite{16} where scalar fields play the role of dark energy, the quintessence model is the most studied one. \par

Modified gravity theories can be constrained by observations. The gravitational wave observation of GW170817
and the corresponding gamma ray burst (GRB 170817 A) put strong constraints on the viability of dark energy models in modified gravity \cite{16a,16b}. The viability of the scalar-tensor theories of gravity and Milgrom’s modified Newtonian daynamics (MOND)
with this new observations has been discussed in \cite{16c,16d}. 

Gauss-Bonnet gravity is a modified gravity theory lives in high dimensions with its action given by \cite{17,18}
\begin{equation}
S=\int d^Dx \sqrt{-g} \left(\frac{R-2\Lambda}{16\pi G}+\alpha ~{\cal G} \right)+S_m
\end{equation}
Where ${\cal G}$ is the Gauss-Bonnet invariant defined as 
\begin{equation}
{\cal G}=R^{\mu\nu\alpha\beta}R_{\mu\nu\alpha\beta} - 4R^{\mu\nu}R_{\mu\nu} + R^2
\end{equation}
$\alpha$ is a constant, $S_m$ is the matter action, $R$ is the Ricci scalar and $G$ is Newton's gravitational constant. A new Gauss-Bonnet 4D theory has been introduced in \cite{19} by rescaling $\alpha \rightarrow \frac{\alpha}{D-4}$. Observational constraints have been placed on the free parameter $\alpha$ by some groups and here we use $\alpha=(1.2 \pm 5.2)\times 10^{-17}$ \cite{18,20}. \newline The modified Einstein equations in D dimensions are given by
\begin{equation}
R_{\mu\nu}-\frac{1}{2}g_{\mu\nu}R+\Lambda g_{\mu\nu}+\frac{\alpha}{D-4}(4R_{\mu\alpha\beta\sigma}R_{\nu}^{~\alpha\beta\sigma}-8R_{\mu\alpha\nu\beta}R^{\alpha\beta}-8R_{\mu\alpha}R_{\nu}^{~\alpha}+4RR_{\mu\nu}-g_{\mu\nu} {\cal G})=8\pi G T_{\mu\nu}.
\end{equation}
The modified Friedmann equations for a spatially flat 4D FRW universe can be written as 
\begin{equation} 
\frac{\dot{a}^2}{a^2}+2\alpha \frac{\dot{a}^4}{a^4} = \frac{8\pi G}{3}(\rho_b+\rho_{cdm}+\rho_r)+\frac{\Lambda}{3},  \label{cosm1}
\end{equation}
$\rho_b , \rho_{cdm}$ and $\rho_r$ are the energy densities of baryons, CDM and radiation respectively. The conservation equation is
\begin{equation} 
\dot{\rho_i}+3H(p_i+\rho_i) = 0 .\label{conserv}
\end{equation}
In the current work we assume a time-dependent cosmological constant $\Lambda(t)$. A new cosmological constant model has been introduced in \cite{21} in which the ansatz 
\begin{equation} \label{cosma1}
\Lambda =  \frac{\Lambda_{Pl}}{\left(t/t_{Pl}\right)^2} \propto \frac{1}{t^2}
\end{equation}
 has been proposed. The cosmological constant $\Lambda \propto \frac{1}{t^2}$ starts at the Plank time as $\Lambda_{Pl} = \sim M_{Pl}^2$ and allows the value $\Lambda_{0} \sim 10^{-120} M_{Pl}^2$ for the present epoch. 
Some other general models for the varying cosmological constant have been introduced (see \cite{20b,20c,20cc} and references therein. The following ansatz for $\Lambda(H)$, with $H$ the Hubble constant, has been first introduced in \cite{20d}
\begin{equation} \label{vary2}
\Lambda(H)= \lambda +\alpha H + 3 \beta H^2 
\end{equation}
where $\lambda$, $\alpha$ and $\beta$ are real constants. It was found in \cite{20b, 20f, 20g, 20h} that the case for $\lambda=0$ doesn't fit with observational data while the case of $\lambda \neq 0$ behaves like $\Lambda$CDM model at late-time. Some other models for varying $\Lambda$ in terms of $H$ are \cite{20b}
\begin{eqnarray} 
\Lambda(H)= \beta H +3H^2 + \delta H^n,\,\,\,\,\,\, n \in R-\left\{0,1\right\} \\
\Lambda(H, \dot{H}, \ddot{H})=\alpha+\beta H+\delta H^2+ \mu \dot{H}+\nu \ddot{H}.
\end{eqnarray}

In the current work, we first follow \cite{21} and use $\Lambda(t) = \frac{C}{t^2}$ as an ansatz for the varying $\Lambda$ where $C$ is a constant. This allows the cosmological constant to have a very tiny positive value at the current epoch and late-times as suggested by observations \cite{1,1a} (Figure 2h). We then re-investigate the same model using the ansatz (\ref{vary2}): $\Lambda(H)= \lambda +\alpha H + 3 \beta H^2 $ and compare the results of the two models.

\section{Cosmological solutions and the ad hoc approach}

In the current work, we probe the cosmology of 4D Gauss-Bonnet gravity through an ad hoc approach by using a given scale factor derived from the desired cosmic behavior \cite{ellis} which is inverse of the usual viewpoint. The choice of the scale factor is restricted by satisfying some observational requirements mainly for the deceleration and jerk parameters. The evolution of these two parameters should lead to a cosmic deceleration-acceleration transit along with a flat $\Lambda CDM$ model at late-time. This ad hoc approach for the scale factor and scalar fields has been used extensively in the literature to probe the cosmology of several gravity theories \cite{ellis,el2,sen,senta, sent,sent11,sz,sch,ric}.
\subsection{Model 1}
The following hyperbolic ansatz leads to a desired behavior of the deceleration and jerk parameters for $0<n<1$ \cite{nasrtarek}
\begin{equation} \label{ansatz}
a(t)= A \sinh^{\frac{1}{n}}(\eta t)
\end{equation}
The expressions for the deceleration parameter $q$, the jerk parameter $j$, the cosmic pressure $p$, the energy density $\rho$ are 
\begin{equation} \label{q1}
q=-\frac{\ddot{a}a}{\dot{a}^2}=\frac{-\cosh^2(\eta t)+n}{\cosh^2(\eta t)},~~~ j=\frac{\dddot{a}}{aH^3}= 1+\frac{2n^2-3n}{\cosh^2(\eta t)}~~~~~~~~~~~~~~~~~~~~~
\end{equation}
\begin{eqnarray} 
\rho=-\frac{-6\alpha g^4 \eta^{4}t^2-3g^2\eta^2 t^2n^2f^2+Cn^4f^4}{t^2n^4f^4}~~~~~~~~~~~~~~~~~~~~~~~~~~~~~~~~~~~~~~\\   \label{rho}
p=-\frac{1}{g\eta t^3n^4f^4}\left( 24 f^2g^3 n\alpha t^3n^5+6f^4g n^3\eta^3t^3-6f^2g^3n^3\eta^3t^3+2n^5Cf^5 \right. \\   \nonumber
\left.-24n\alpha t^3\eta^5g^5+18\alpha t^3n^5g^5+9n^2t^3\eta^3f^2g^3-3Cn^4\eta t f^4g \right)
\end{eqnarray}
Where $f=\sinh(\eta t)$ and $g=\cosh(\eta t)$. The plots of $q$ and $j$ are shown in figure 1 for different values of $n$. We will consider $n=2$ where $q$ changes sign from positive to negative, and $j$ tends to $+1$ at late-time as expected for a flat $\Lambda$CDM universe.

The figure shows that $q$ varies in the range from $-1$ to $+1$ where cosmic transit should happen at $\ddot{a}=0$ ( $q=0$ ). Here we get $t_{q=0}=\frac{1}{2 \eta}\ln(3+2\sqrt{2})$ which gives $t\approx 0.88$ for $\eta=1$. The Hubble parameter plotted as a function of the redshift in Fig.1(c). Cosmic pressure $p$ changes sign from positive to negative in corresponding to the sign flipping of the deceleration parameter $q$. Taking into account the Dark Energy assumption as a component of negative pressure that gives the effect of a repulsive gravity, cosmic pressure should be positive in the early-time decelerating era and negative in the late-time decelerating era \cite{cycl}. After expressing the equation of state parameter $\omega$ in terms of the redshif $z$ using $a=\frac{1}{1+z}$, we find that $\omega(z)=-1$ at the current epoch where $z=0$ as suggested by observations (Fig.1(f)). The evolution of $\omega$ against cosmic time (Fig.1(g)) shows a Quintessence dominated universe where $ -1 \leq \omega \leq 0.4$ with no crossing to the phantom divide line $\omega=-1$. The no-go theorem introduced in \cite{nogo} forbids $\omega$ of a single perfect fluid in FRW universe to cross the phantom divide line, this no-go theorem doesn't hold in higher dimensions \cite{extra}. 

\subsubsection{Energy and Causality Conditions}
The classical linear energy conditions \cite{ec11,ec12} and the new nonlinear energy conditions (ECs) \cite{ec,FEC1, FEC2, detec} have been plotted in Fig.1 (h) \& (i). The classical linear ECs (`` the null $\rho + p\geq 0$; weak $\rho \geq 0$, $\rho + p\geq 0$; strong $\rho + 3p\geq 0$ and dominant $\rho \geq \left|p\right|$ energy conditions '' ) should be replaced by other nonlinear ECs when semiclassical quantum effects are taken into account \cite{ec, detec}. Here we consider (i) The flux EC (FEC): $\rho^2 \geq p_i^2$ \cite{FEC1, FEC2}, first presented in \cite{FEC1}. (ii) The determinant EC (DETEC): $ \rho . \Pi p_i \geq 0$ \cite{detec}. (ii) The trace-of-square EC (TOSEC): $\rho^2 + \sum p_i^2 \geq 0$ \cite{detec}. According to the strong energy condition (SEC), gravity should always be attractive. But this `highly restrictive' condition fails when describing the current cosmic accelerated epoch and during inflation \cite{ec3,ec4}. Here we have a change of sign in cosmic pressure from positive to negative and, consequently, the SEC is not expected to be satisfied as indicated in Fig. 1(h). The null energy condition (NEC) and the dominant energy condition are satisfied all the time. The NEC is the most fundamental of the ECs and on which many key results are based such as the singularity theorems \cite{necvb}. Violation of NEC automatically implies the violation of all other point-wise energy conditions. We have found that the condition $0 \leq \frac{dp}{d\rho} \leq 1$ is not valid and so the sound speed causality condition is not satisfied.

\subsection{Re-investigating Model 1 with $\Lambda(H)= \lambda +\alpha H + 3 \beta H^2$ }
Using (\ref{vary2}) along with (\ref{ansatz}) into (\ref{cosm1}) and (\ref{conserv}) we obtain the following expressions for cosmic pressure and energy density
\begin{eqnarray} 
p&=&\frac{\eta \alpha}{3} \tanh(\eta t)+\frac{2\eta^2 \beta}{n} -\frac{2\eta^2 }{n} -\coth^2(\eta t)( \frac{8\eta^4 \alpha}{n^3} +\lambda+\frac{3\eta^2 \beta}{n^2}-\frac{2\eta^2 \beta}{n}-\frac{\eta^2}{n^2} +\frac{2\eta^2 }{n})\\ \nonumber
&+&\coth^4(\eta t)(\frac{8\eta^4 \alpha}{n^3} -\frac{6\eta^4 \alpha}{n^4})-\coth(\eta t)(\frac{\eta \alpha}{3} -\frac{\eta \alpha}{n})\\ \nonumber
\rho&=&-\lambda-\frac{\eta \alpha}{n}\coth(\eta t)-\frac{3\eta^2}{n^2}\coth^2(\eta t)(1-\beta)+ \frac{6\eta^4 \alpha}{n^4}\coth^4(\eta t)
\end{eqnarray}
The behavior of $p$, $\rho$, $\omega$, energy conditions and the sound speed has been plotted in fig(\ref{tap221})

\begin{figure}[H] \label{tap1}
  \centering             
  \subfigure[$q$]{\label{F1}\includegraphics[width=0.29\textwidth]{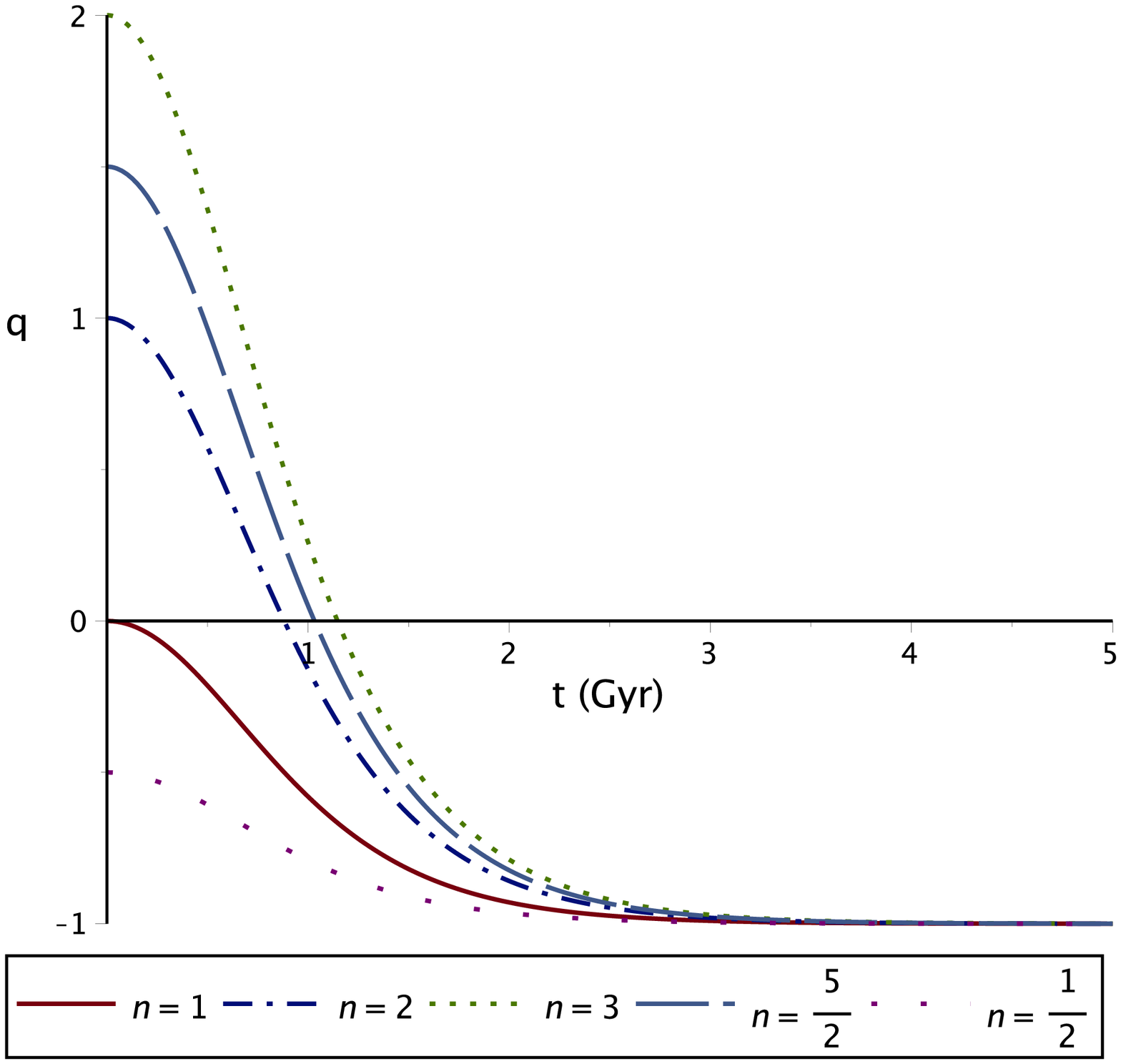}} 
	\subfigure[$j$]{\label{F2}\includegraphics[width=0.29\textwidth]{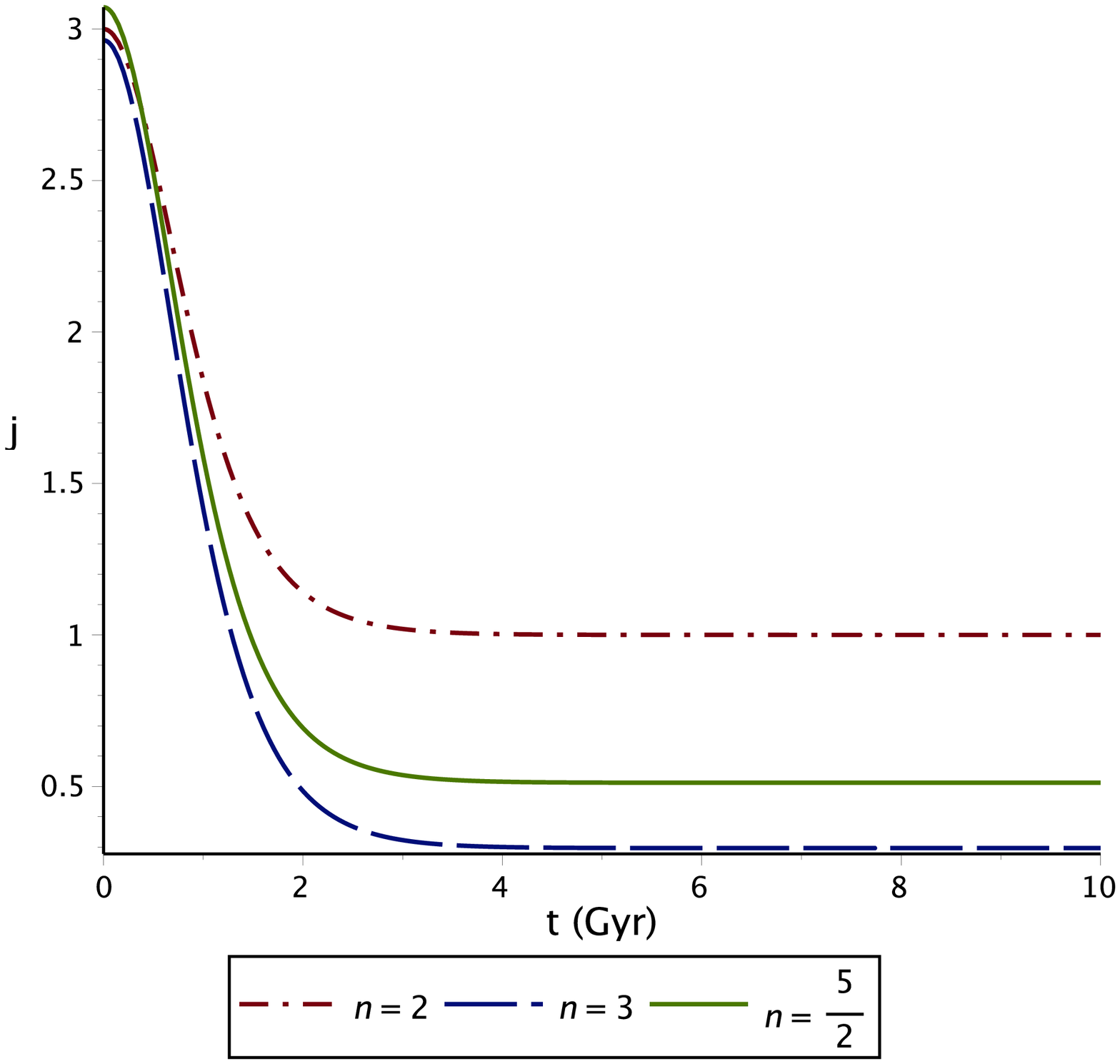}} 
	  \subfigure[$H(z)$]{\label{F3}\includegraphics[width=0.29\textwidth]{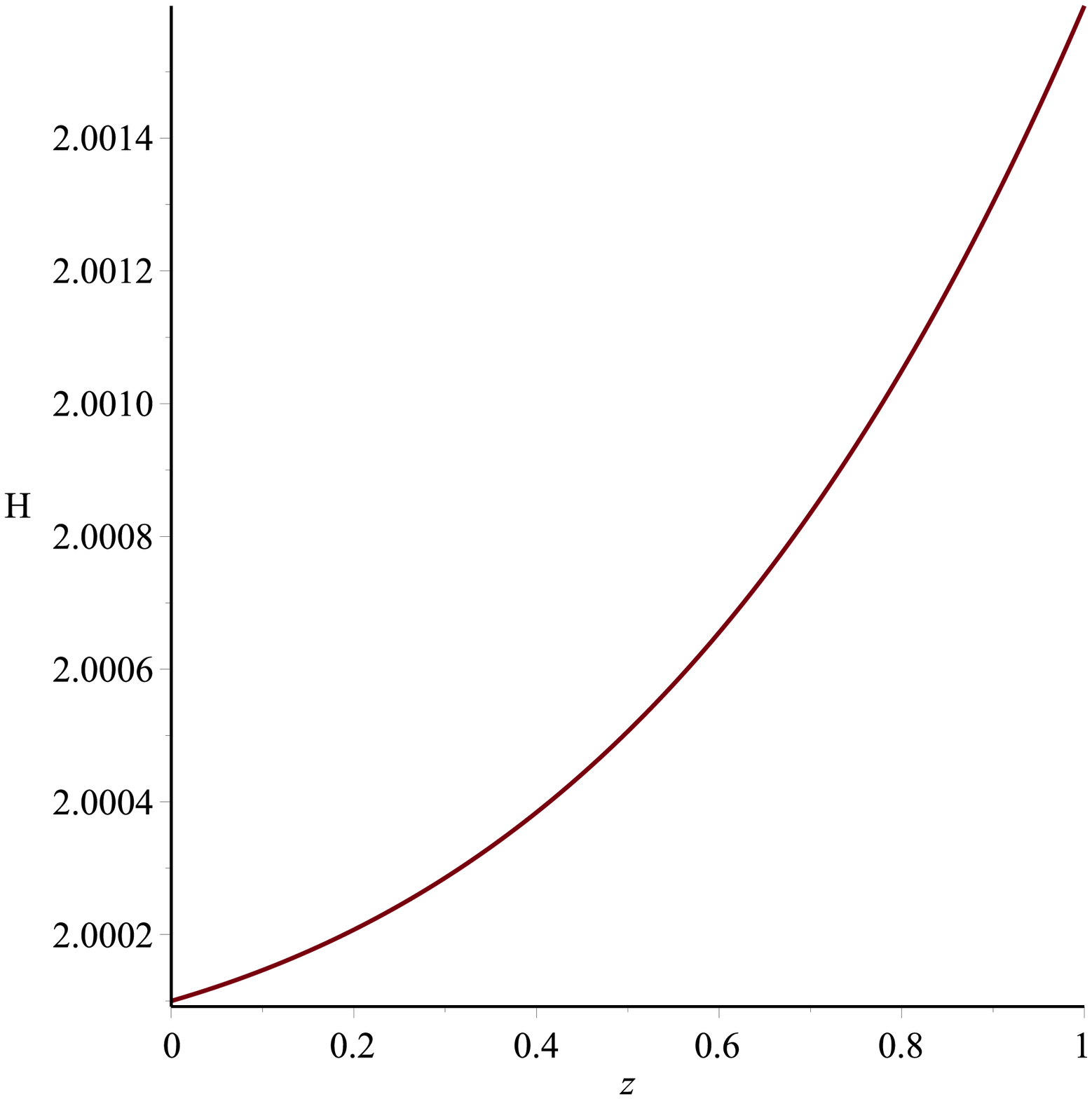}} \\
			  \subfigure[$p$]{\label{F4}\includegraphics[width=0.29 \textwidth]{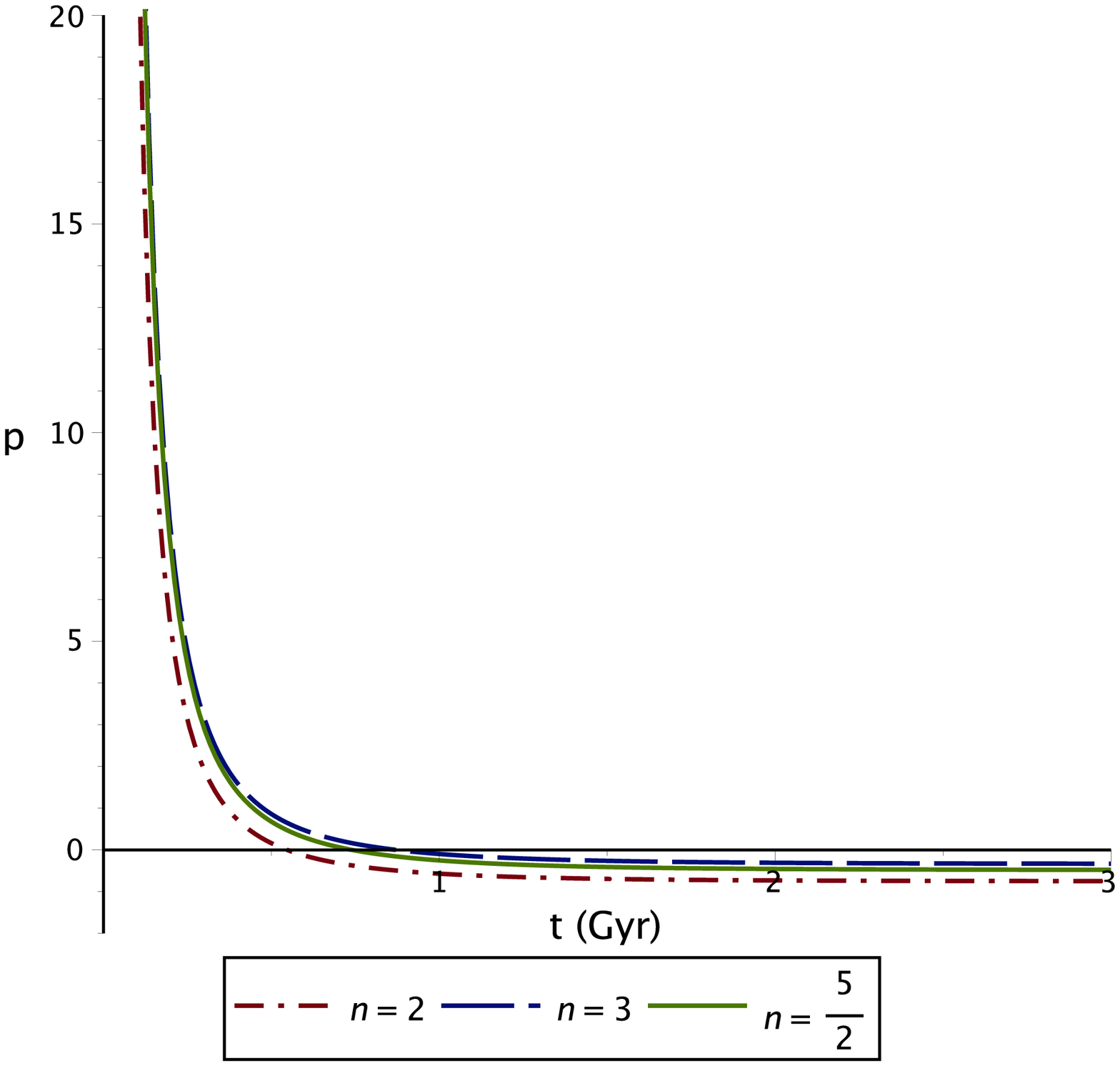}}
				\subfigure[$\rho$]{\label{F5}\includegraphics[width=0.29 \textwidth]{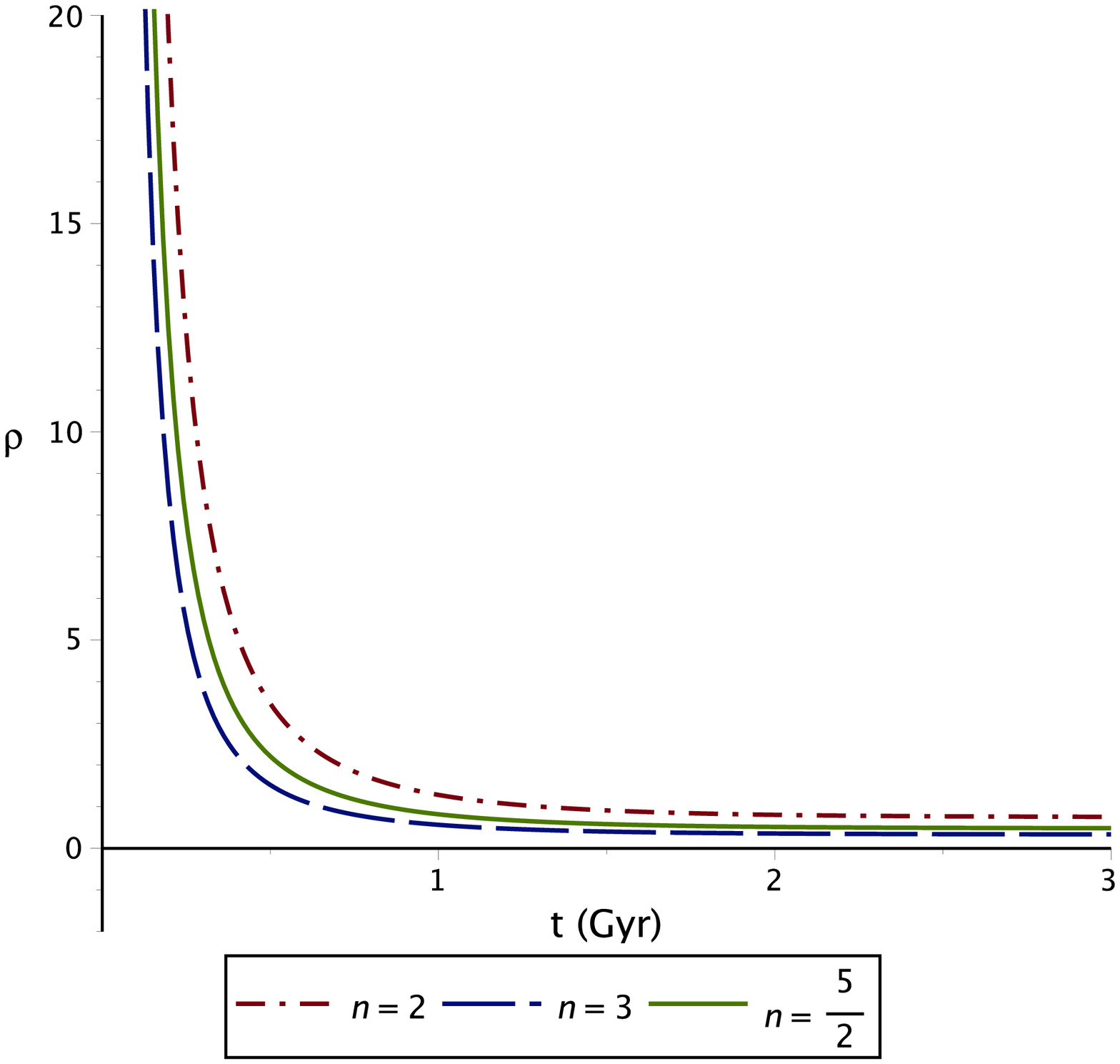}}
	\subfigure[$\omega(z)$]{\label{F6}\includegraphics[width=0.29\textwidth]{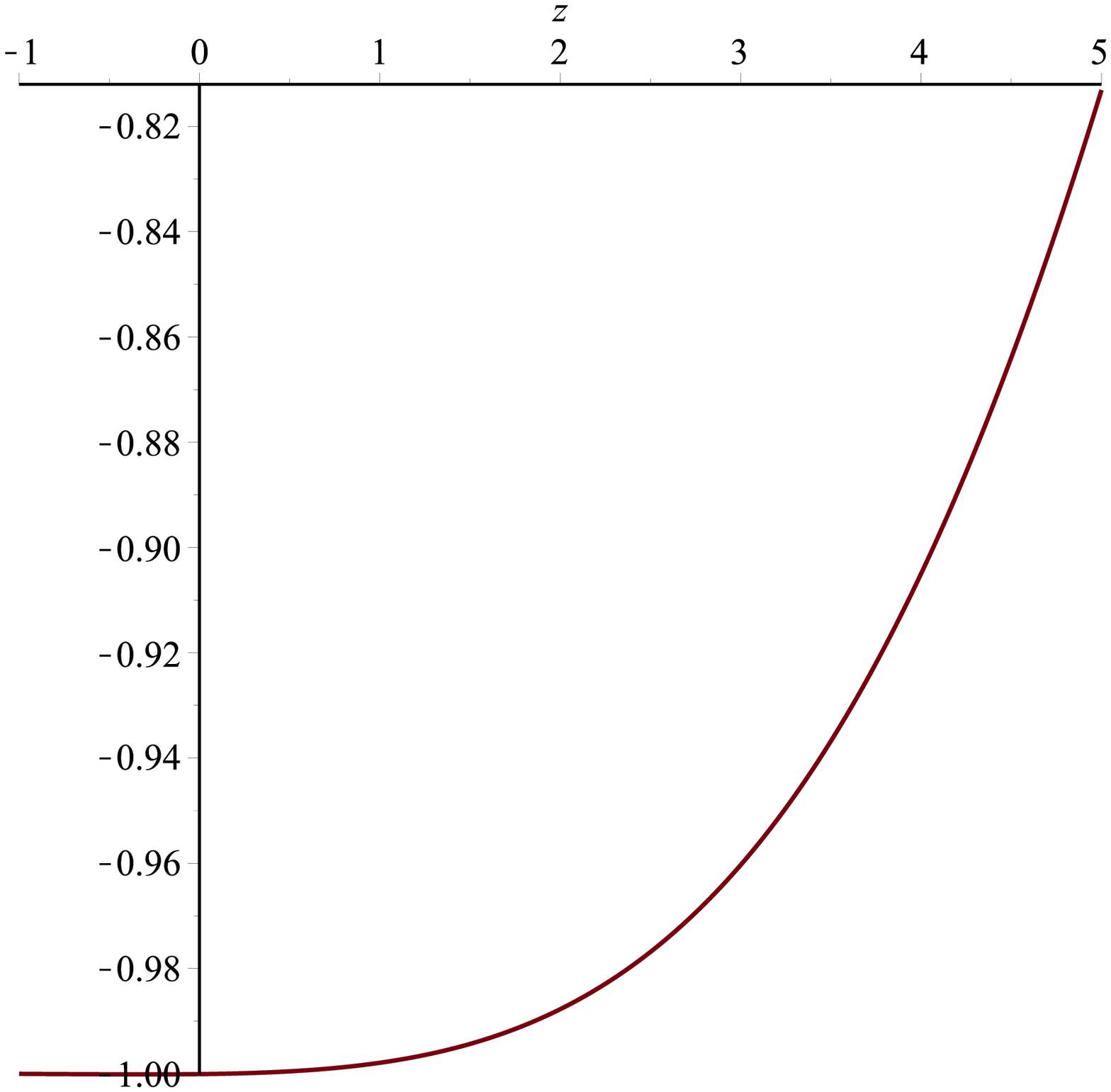}} \\
	\subfigure[$\omega(t)$]{\label{F7}\includegraphics[width=0.29\textwidth]{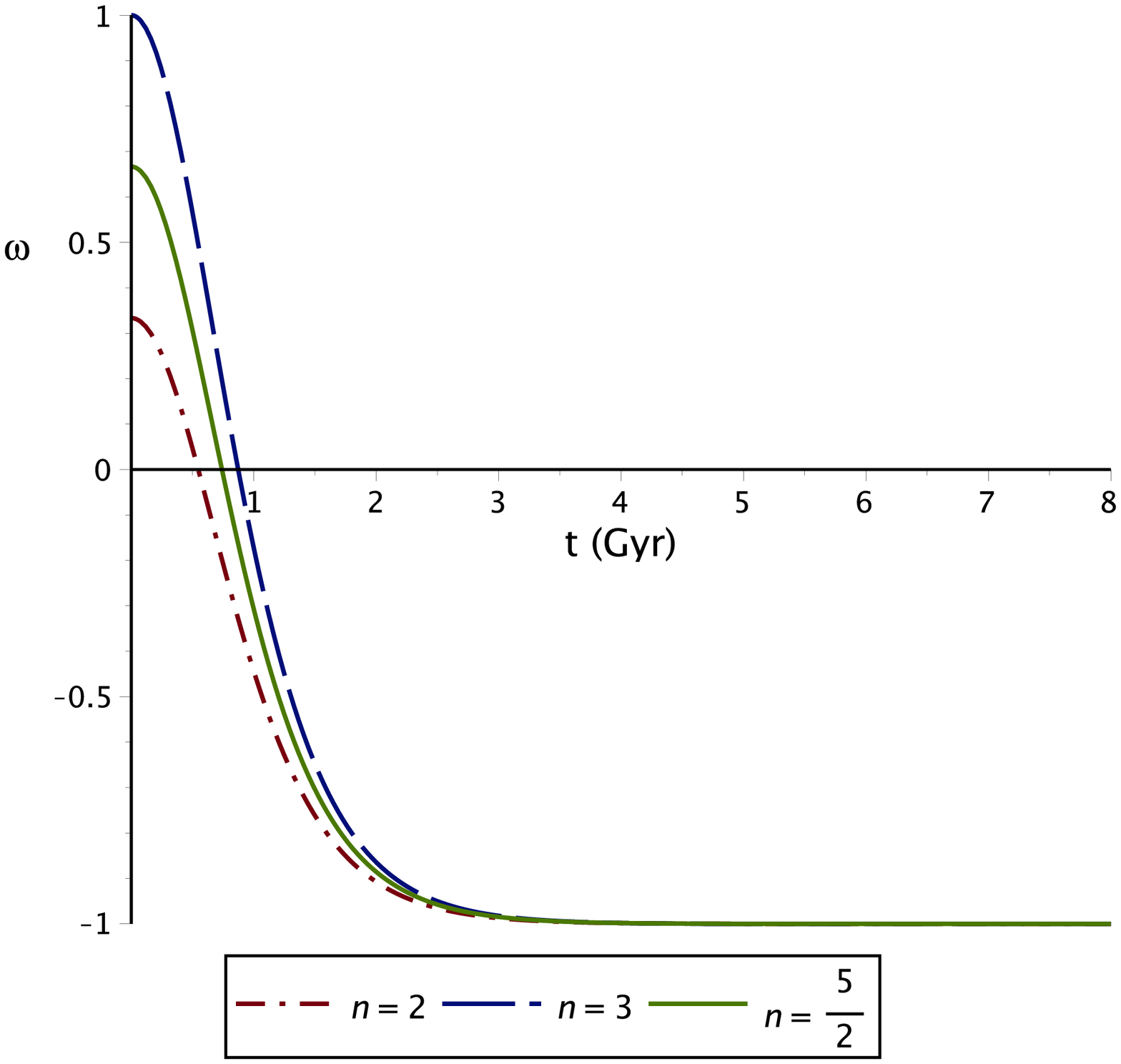}} 
		\subfigure[$ECs)$]{\label{F8}\includegraphics[width=0.29\textwidth]{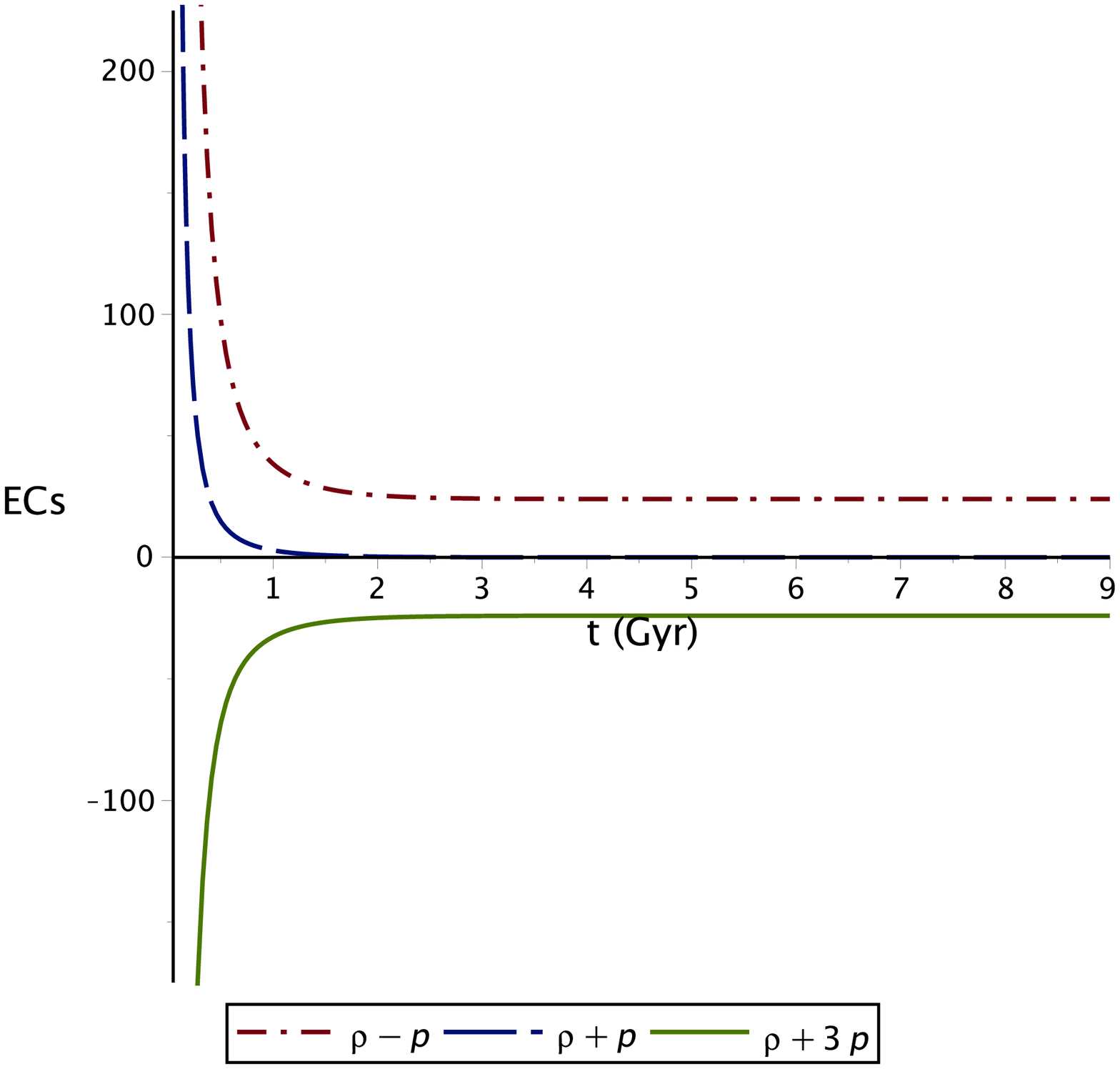}}
			\subfigure[$nlECs$]{\label{F9}\includegraphics[width=0.29\textwidth]{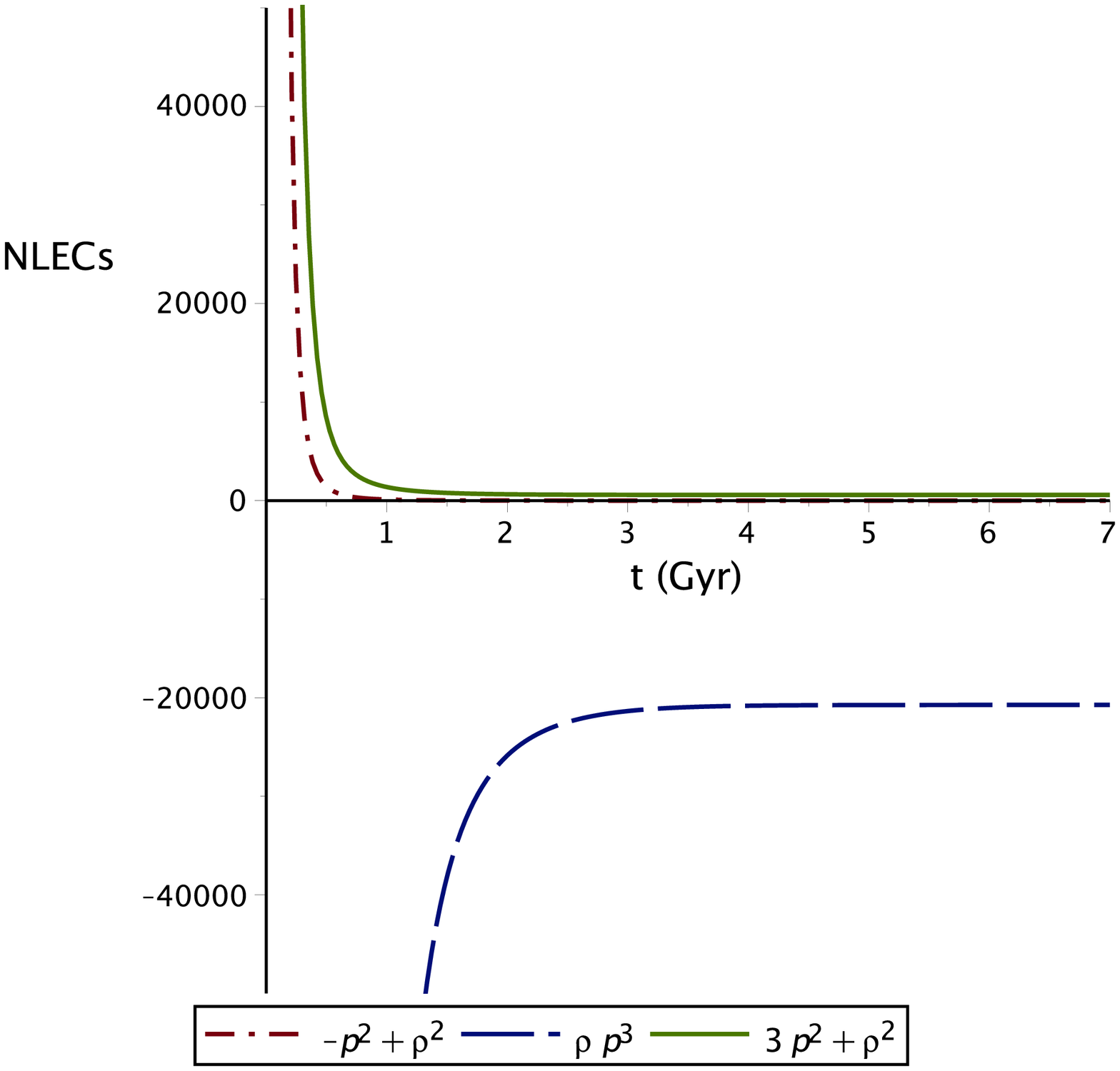}}
  \caption{The hyperbolic solution using $\Lambda(t)=\frac{C}{t^2}$: (a) The deceleration parameter $q$ changes sign from positive to negative. (b) The jerk parameter $j=1$ at late-times where the model tends to a flat $\Lambda$CDM model. (c) The variation of Hubble parameter against redshift. (d) Cosmic pressure goes from positive to negative. (e) The evolution of energy density. (f) EoS parameter $\omega =-1$ at $z=0$. (g) The EoS parameter $\omega(t)$ tends to $-1$ at late-time with no crossing to the phantom divide line . (h) and (g) shows linear and nonlinear energy conditions. The model fails to pass the sound speed causality condition. Here $\alpha=7.2\times 10^{-17}$, $A=0.1$, $C=0.01$ and $\eta=1$.}
  \label{fig:1}
\end{figure}

\begin{figure}[H] \label{tap221}
  \centering             
  \subfigure[$\rho$]{\label{et1}\includegraphics[width=0.29\textwidth]{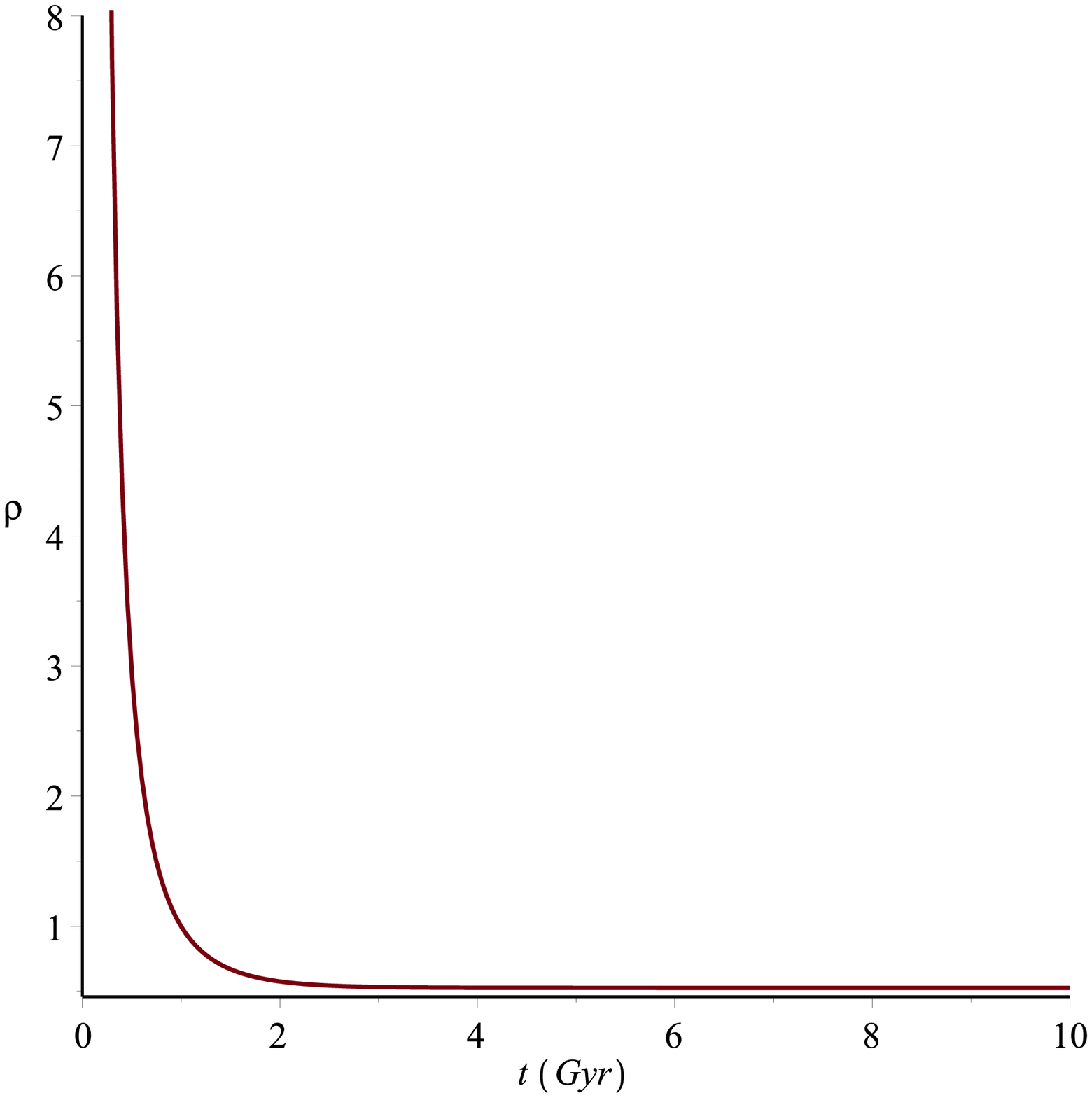}} 
	\subfigure[$p$]{\label{et2}\includegraphics[width=0.29\textwidth]{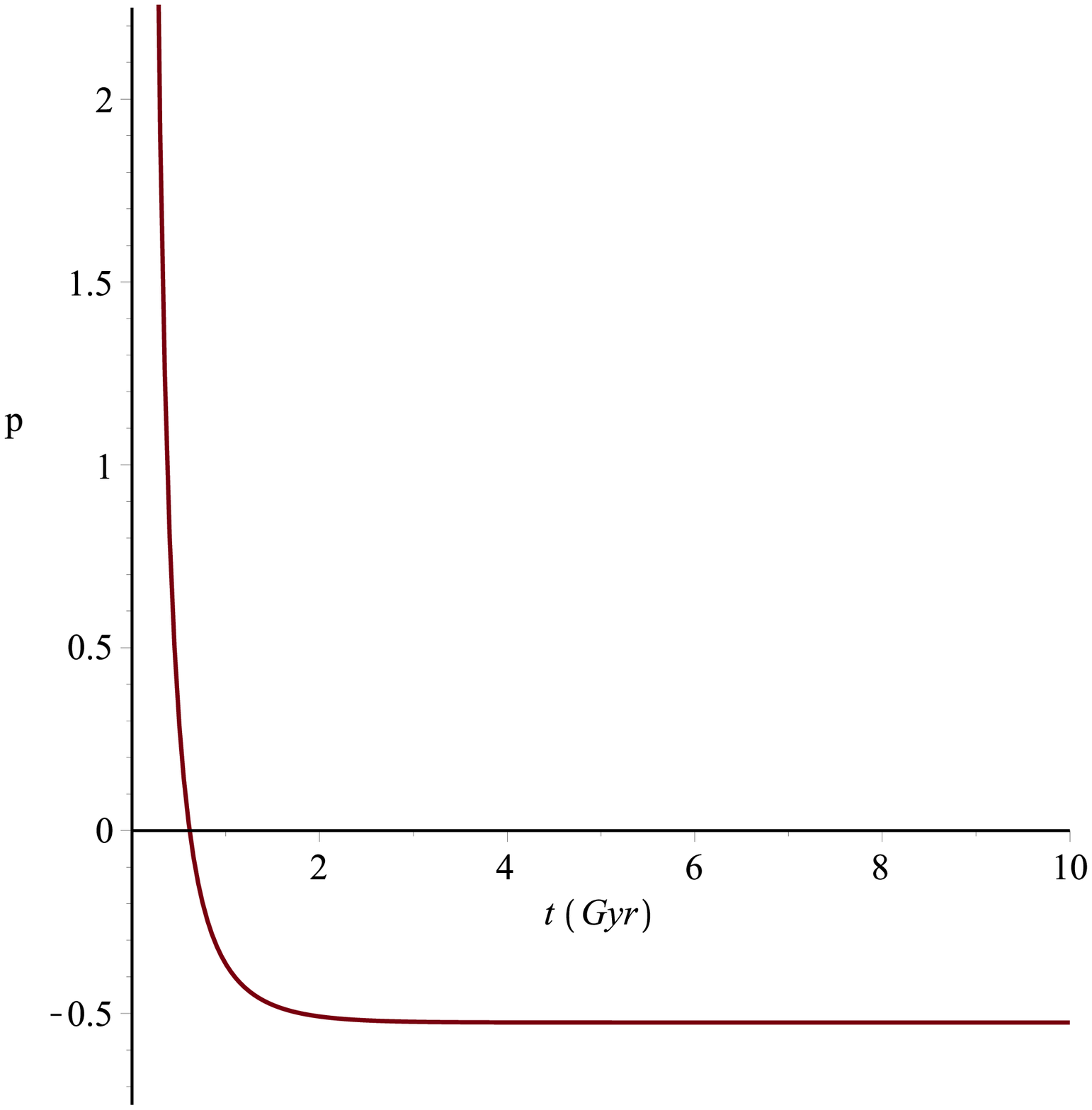}} 
	  \subfigure[$\omega(t)$]{\label{et3}\includegraphics[width=0.29\textwidth]{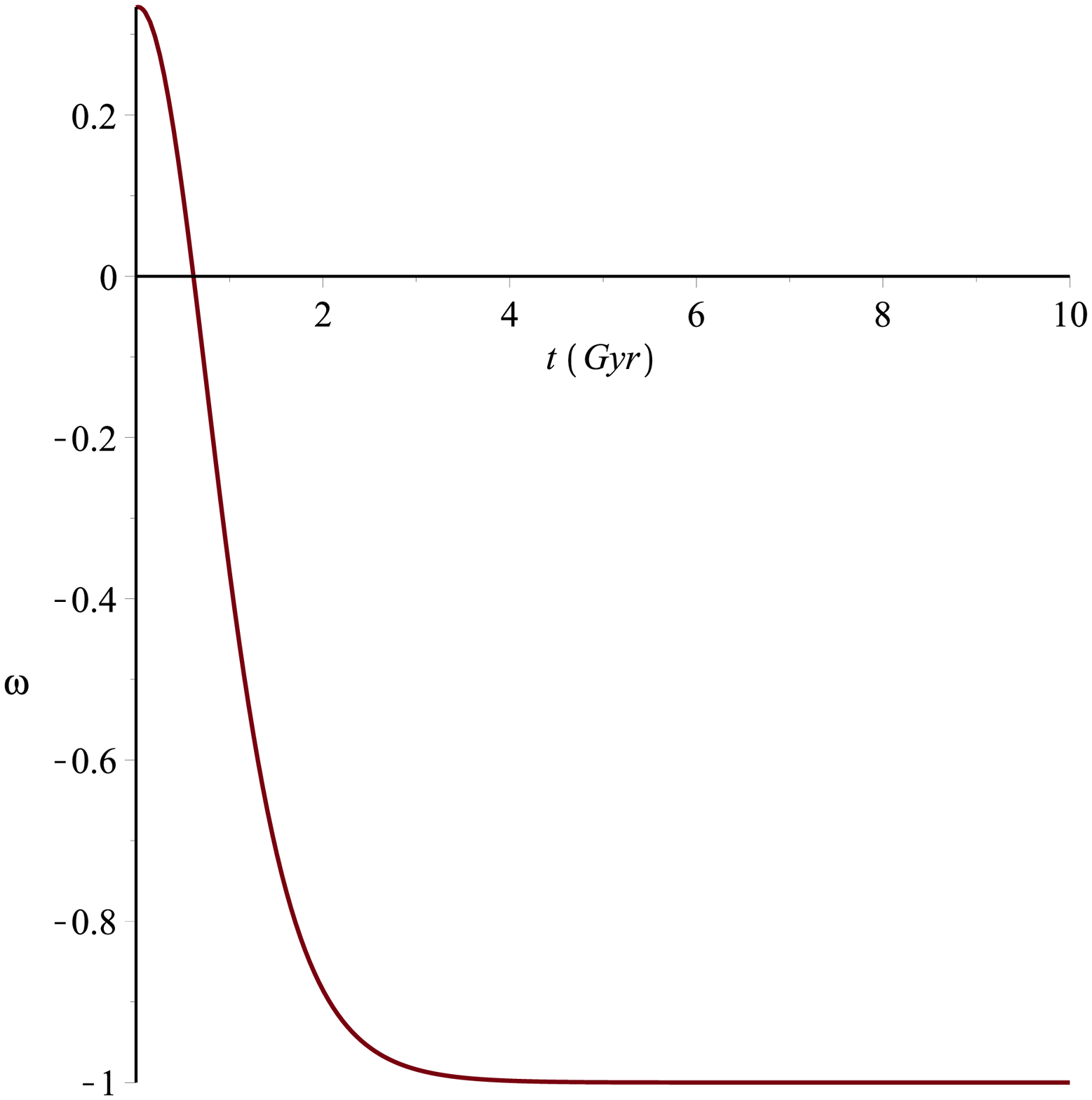}} \\
			  \subfigure[$ECs$]{\label{et4}\includegraphics[width=0.29 \textwidth]{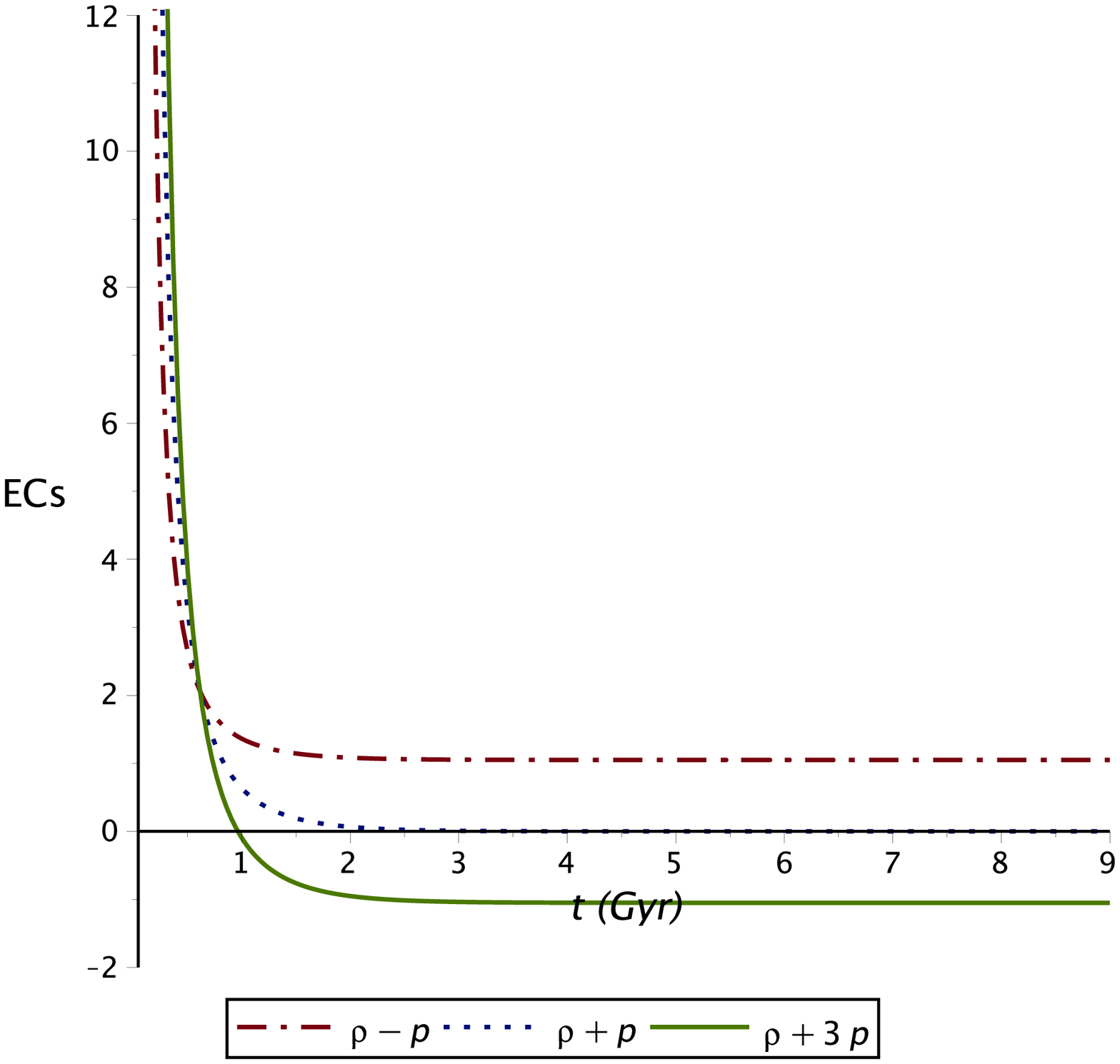}}
				\subfigure[$nlECs$]{\label{et5}\includegraphics[width=0.29 \textwidth]{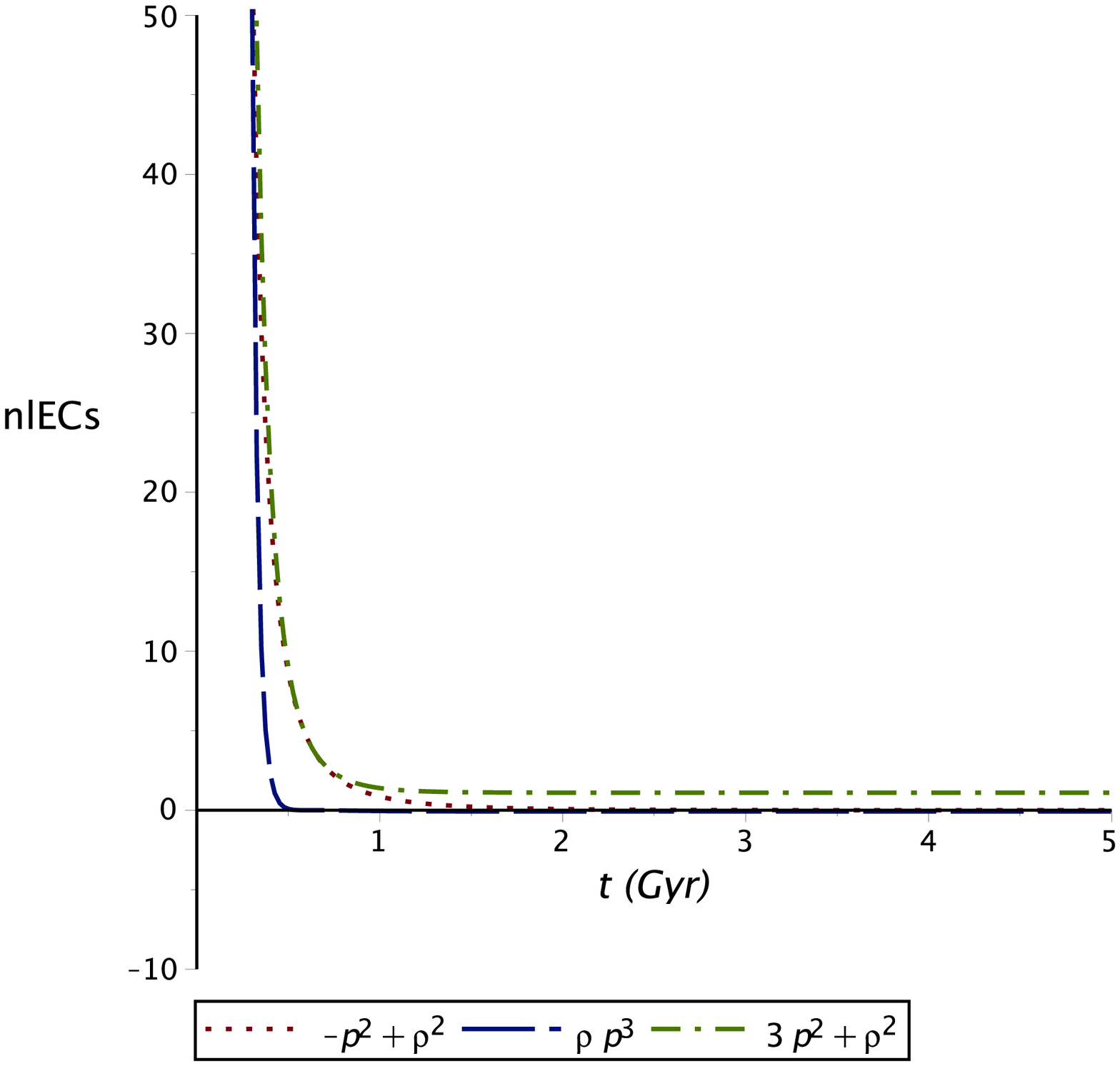}}
				\subfigure[$v_s=\frac{dp}{d\rho}$]{\label{et6}\includegraphics[width=0.29 \textwidth]{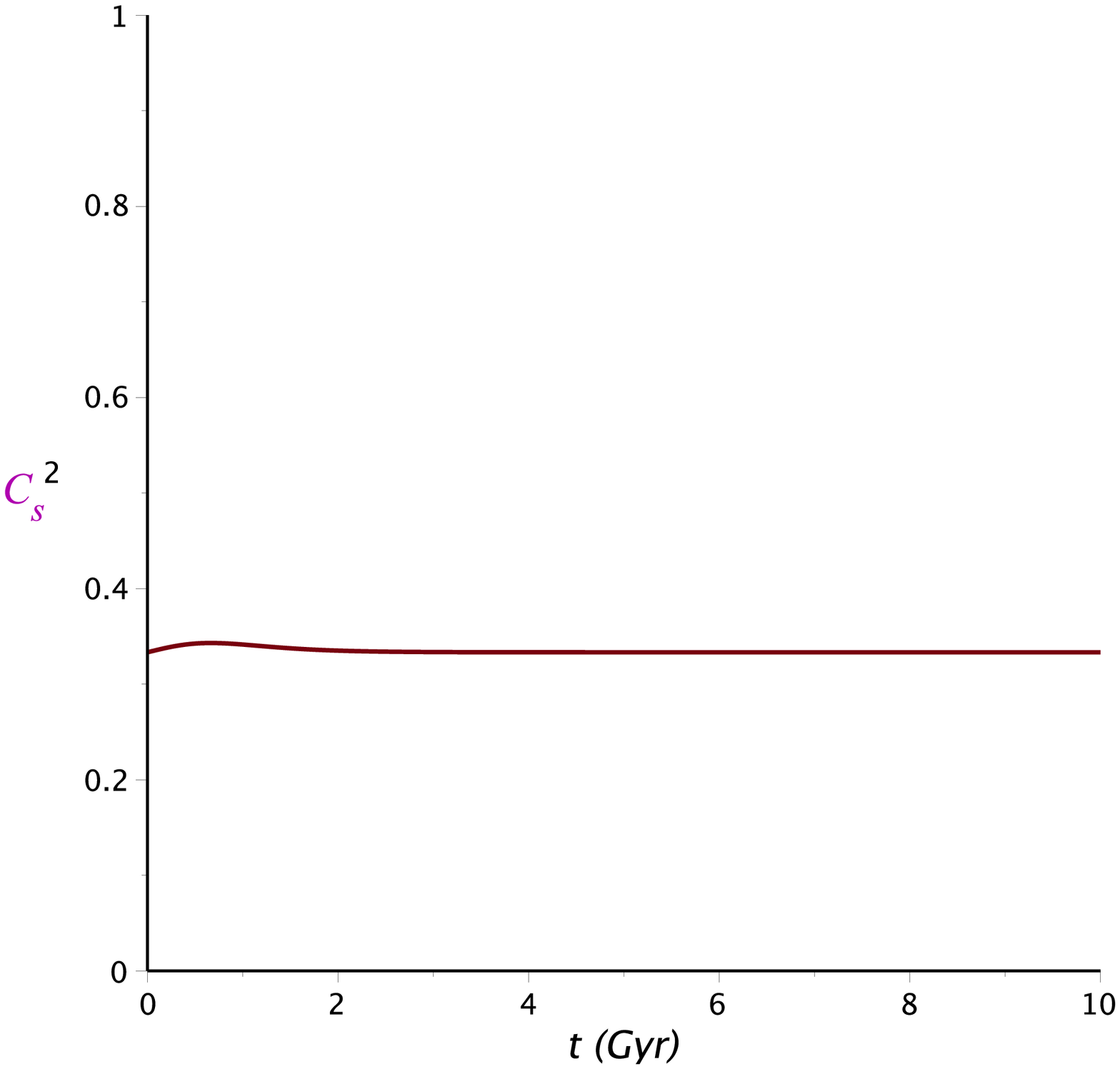}}
  \caption{The hyperbolic solution using $\Lambda(H)= \lambda +\alpha H + 3 \beta H^2 $:  (a) Cosmic pressure goes from positive to negative. (b) The evolution of energy density. (c) $\omega(t)$ tends to $-1$ at late-time with no crossing to the phantom divide line . (d) and (e) shows linear and nonlinear energy conditions, all the nonlinear ECs are satisfied for this case. (f)The sound speed causality condition $0 \leq C^2_s=\frac{dp}{d\rho}\leq 1$ is satisfied for this choice of $\Lambda(H)$ in contrast to the first choice of $\Lambda(t)=\frac{C}{t^2}$ . Here $\alpha=7.2\times 10^{-17}$, $A=0.1$, $\lambda=\alpha=\beta=0.1$, $n=2$ and $\eta=1$.}
\end{figure}

\section{Model 2 (Hybrid solution with $\Lambda(H)= \lambda +\alpha H + 3 \beta H^2$)}
Another ansatz that leads to a desired behavior of the deceleration and jerk parameters (Fig.2 a,b) is the hybrid one given as \cite{hyb1}
\begin{equation} \label{hyb2}
a(t)=a_1 t^{\alpha_1} e^{\beta_1 t},
\end{equation}
where $a_1>0$, $\alpha_1 \geq 0$ and $\beta_1 \geq 0$ are constants. The scale factor (\ref{hyb2}) is a mixture of the power-law and exponential-law cosmologies and a generalization to both of them. The power-law cosmology is obtained for $\beta_1=0$, and the exponential-law cosmology is obtained for $\beta_1=0$. New cosmologies can be explored for $\alpha_1>0$ and $\beta_1>0$. The deceleration and jerk parameter respectively are
\begin{equation}
q(t)=\frac{\alpha_1}{(\beta_1t+\alpha_1)^2}-1 ~~~ ,~~~  j={\frac {{\alpha_1}^{3}+ \left( 3\,\beta\,t-3 \right) {\alpha_1}^{2}+
 \left( 3\,{\beta}^{2}{t}^{2}-3\,\beta\,t+2 \right) \alpha_1+{\beta}^{3}
{t}^{3}}{ \left( \beta\,t+\alpha_1 \right) ^{3}}}
\end{equation}

Using (\ref{hyb2}) in (\ref{cosm1}) and (\ref{conserv}) along with (\ref{vary2}) we get the cosmic pressure and energy density as
\begin{eqnarray} 
\rho&=&\frac{1}{t^4}\left[ (6\alpha \beta_1^4 + 3 \beta_1^2 \beta +3 \beta_1^2 - \alpha \beta_1 - \lambda )t^4 + (24 \alpha \alpha_1 \beta_1^3 -6 \alpha_1\beta_1\beta +6\alpha_1\beta_1 - \alpha_1 \alpha)t^3  \right.  \\  \nonumber
&+&\left.  (36 \alpha \alpha_1^2\beta_1^2-3\alpha_1^2\beta +3\alpha_1^2)t^2 +24\alpha\beta_1\alpha_1^3t +6\alpha \alpha_1^4 \right] \\  \nonumber
p&=&-\frac{1}{3t^4(t\beta_1+\alpha_1)}\left[t^5(18\alpha \beta_1^5 -9\beta_1^3\beta + 9\beta_1^3 - 3\beta_1^2\alpha-3\beta_1\lambda) \right.\\  \nonumber
&+&\left.  t^4(90\alpha \alpha_1\beta_1^4-27\alpha_1\beta_1^2\beta+27\alpha_1\beta_1^2-6\alpha_1\beta_1\alpha-3\alpha_1\lambda)\right.\\  \nonumber
&+&\left.  t^3(180\alpha\alpha_1^2 \beta_1^3-24 \alpha\alpha_1\beta_1^3-27\alpha_1^2\beta_1\beta+27 \alpha_1^2\beta_1-3\alpha_1^2\alpha+6\alpha_1\beta_1\beta-6\alpha_1\beta_1+\alpha_1\alpha) \right.\\  \nonumber
&+&\left.  t^2(180\alpha\alpha_1^3\beta_1^2-27\alpha\alpha_1^2\beta_1^2-9\alpha_1^3\beta+9\alpha_1^3+6\alpha_1^2\beta-6\alpha_1^2)+t(90\alpha \alpha_1^4\beta_1-72\alpha \alpha_1^3\beta_1)\right.\\  \nonumber
&+&\left. 18\alpha\alpha_1^5-24\alpha\alpha_1^4\right]
\end{eqnarray}
The behavior of $\rho$ and $p$ is plotted in Fig.2. The pressure is always positive in contrast to the hyperbolic solution where there is a sign flipping in from positive in the early-time decelerating epoch, to negative in the late-time accelerating epoch. The validity of the linear and non-linear energy conditions at early and late-times are shown in (Fig.2 e and f). We have also found that the sound speed causality condition $0 \leq C^2_s=\frac{dp}{d\rho}\leq 1$ is not satisfied for this model

\begin{figure}[H] 
  \centering             
  \subfigure[$q$]{\label{1}\includegraphics[width=0.29\textwidth]{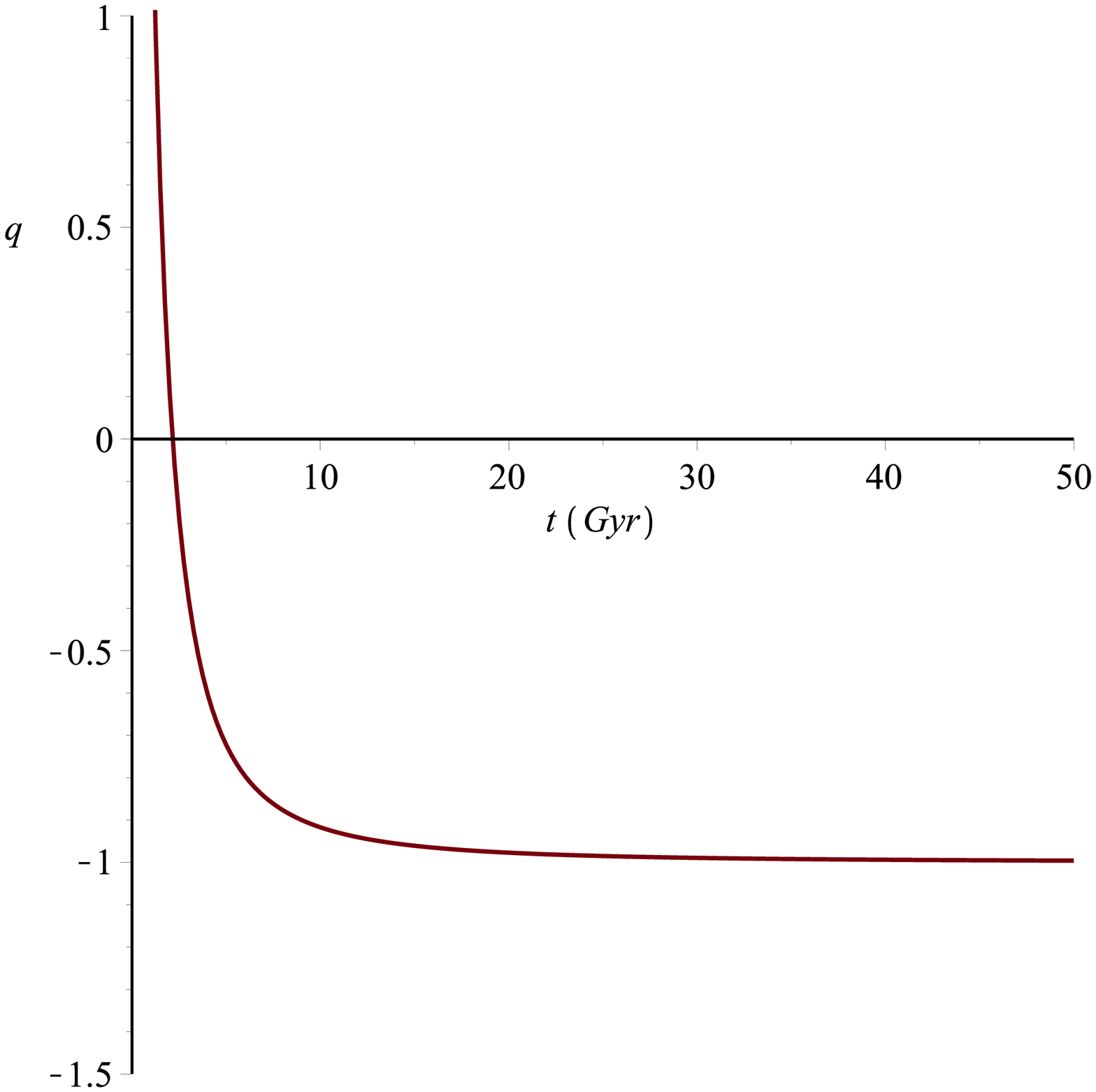}} 
	\subfigure[$j$]{\label{2}\includegraphics[width=0.29\textwidth]{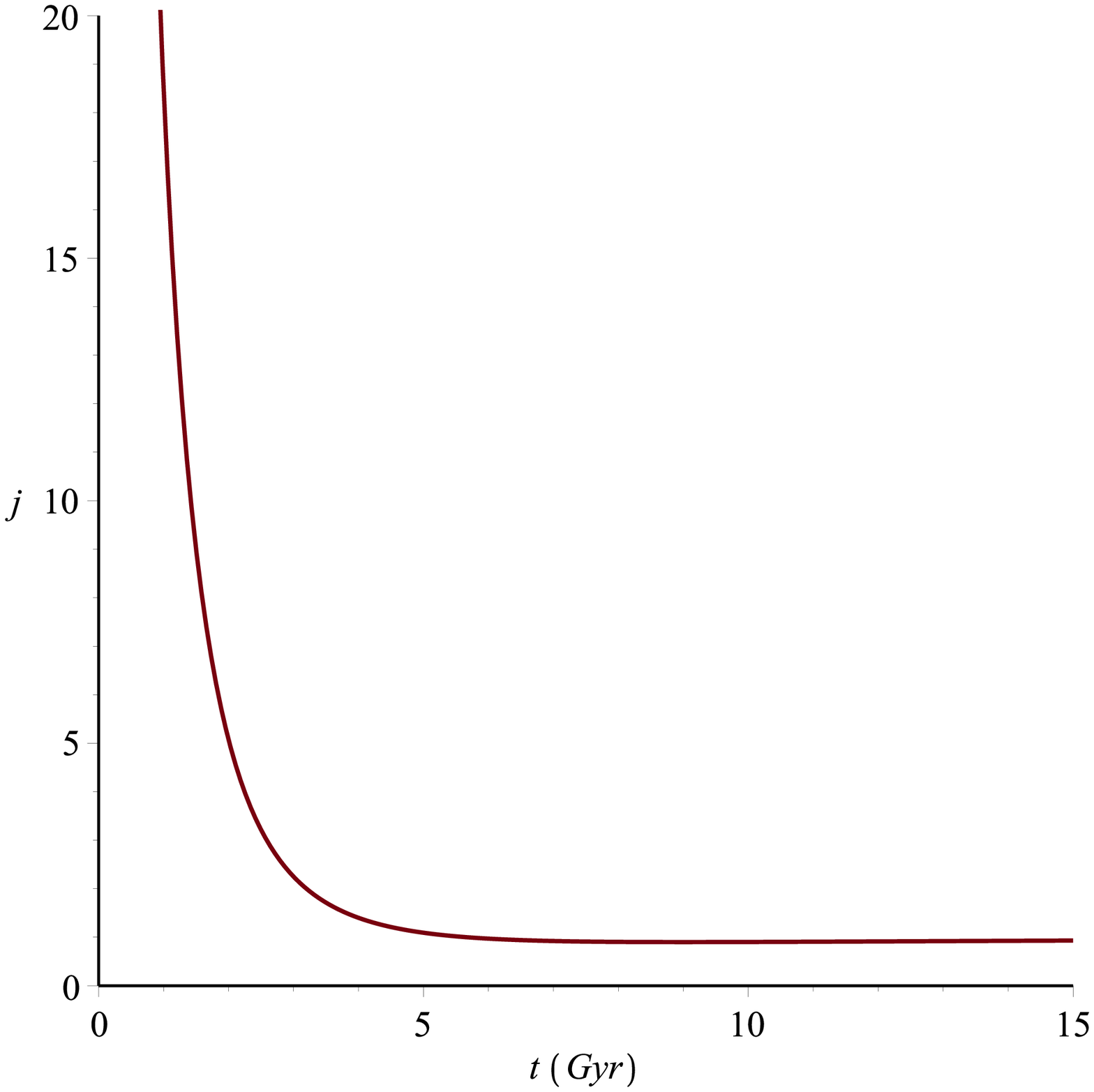}} 
			  \subfigure[$p$]{\label{4}\includegraphics[width=0.29 \textwidth]{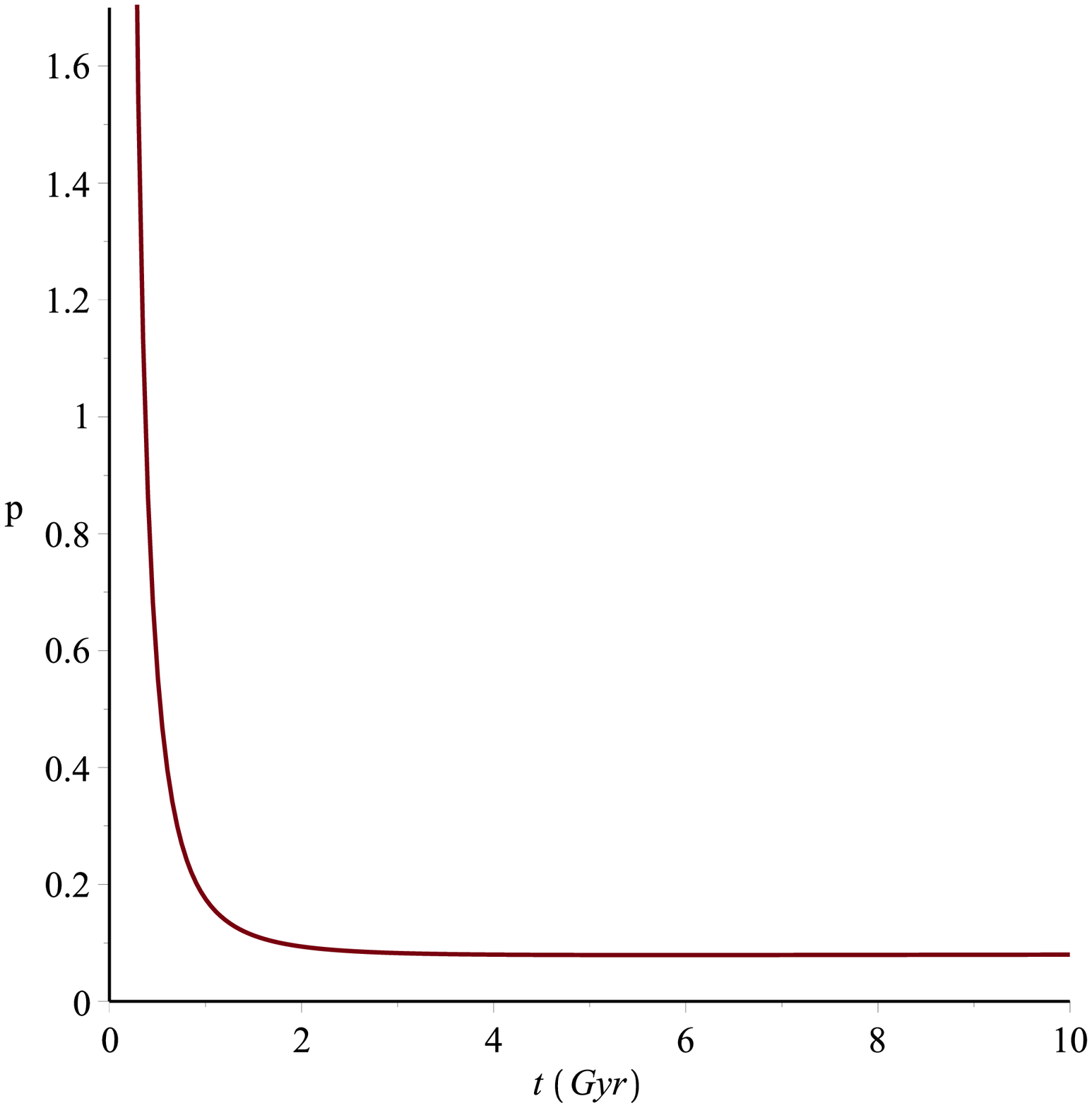}}\\
				\subfigure[$\rho$]{\label{5}\includegraphics[width=0.29 \textwidth]{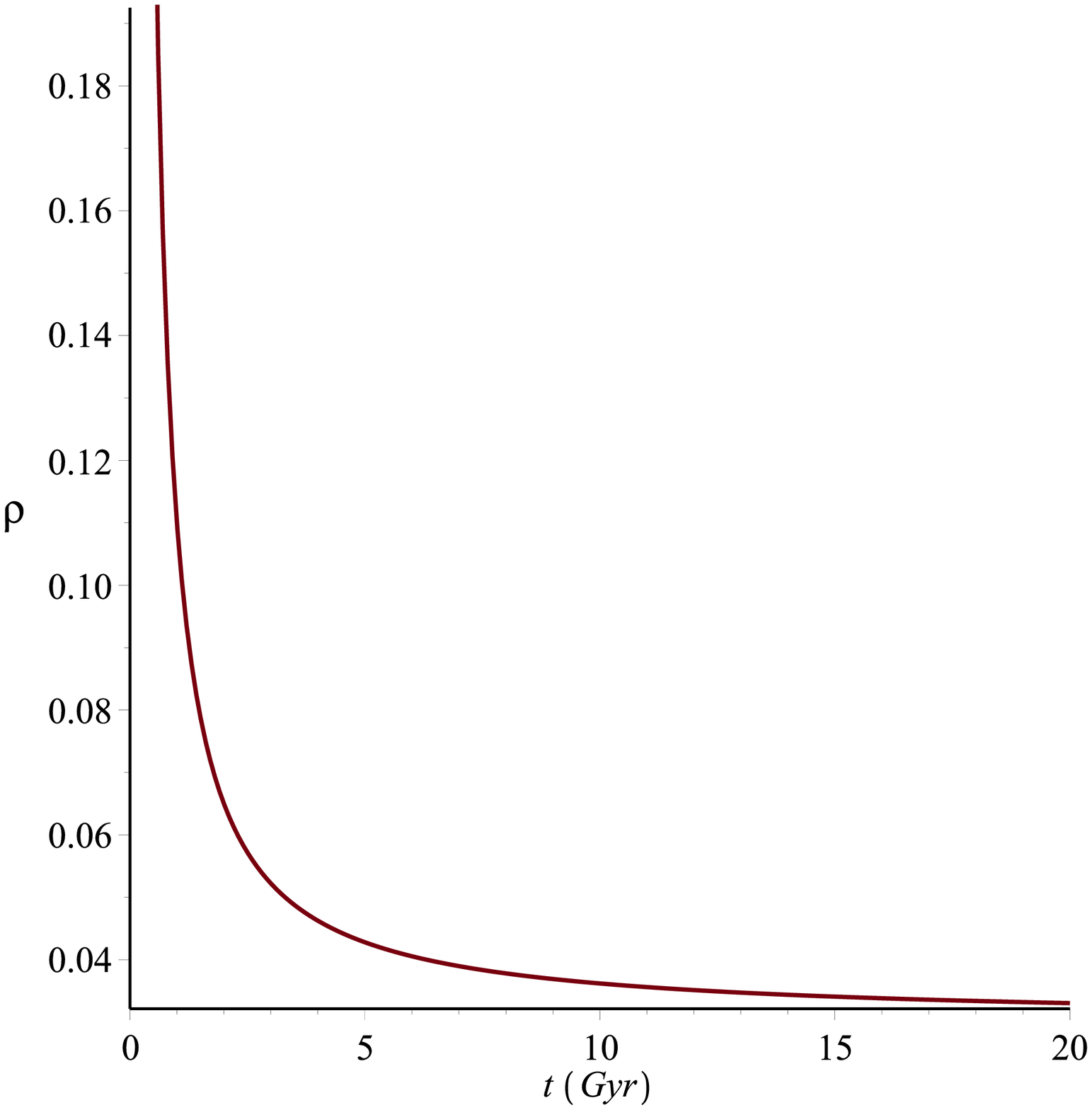}}
		\subfigure[$ECs)$]{\label{8}\includegraphics[width=0.29\textwidth]{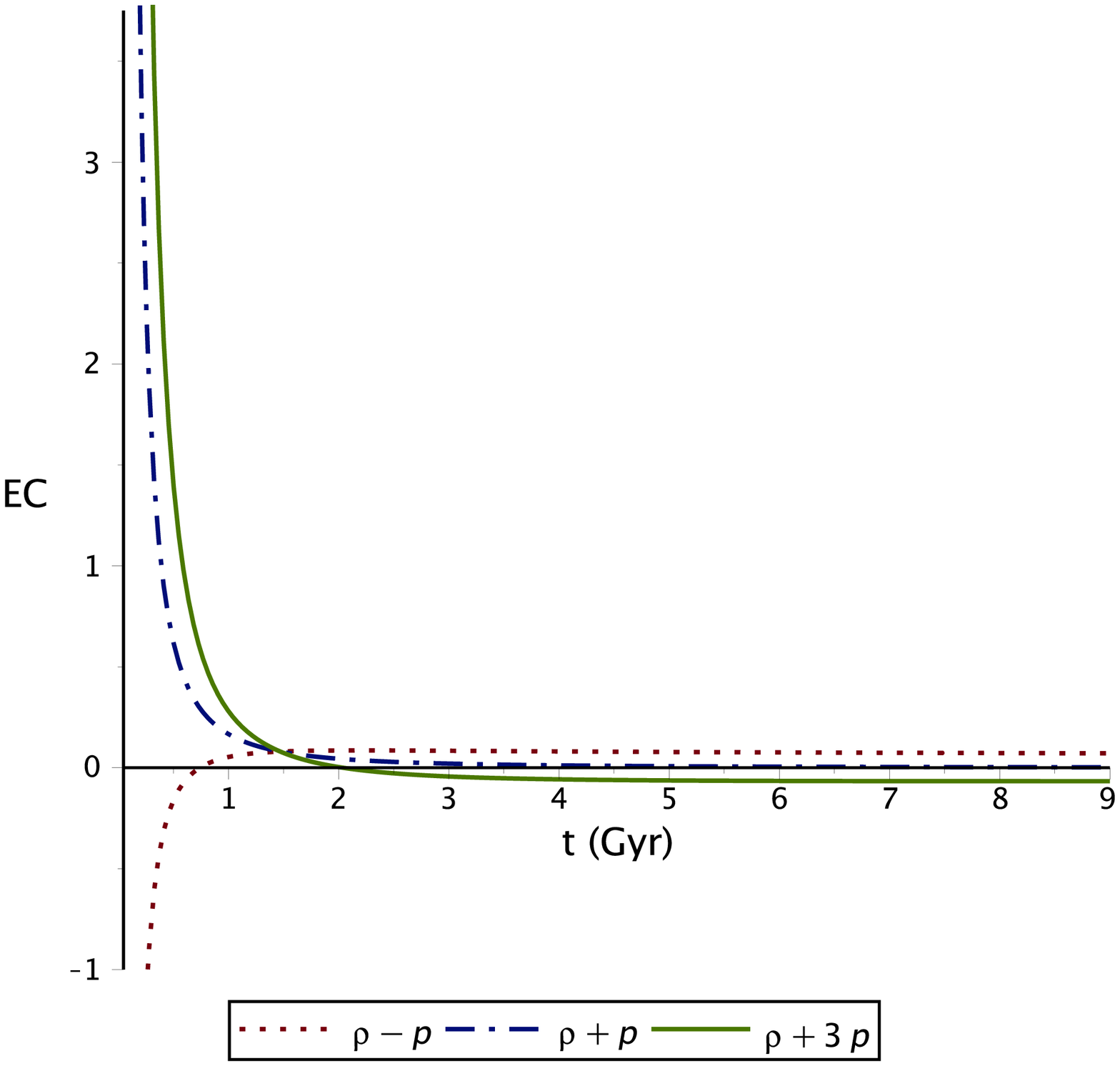}}
			\subfigure[$nlECs$]{\label{9}\includegraphics[width=0.29\textwidth]{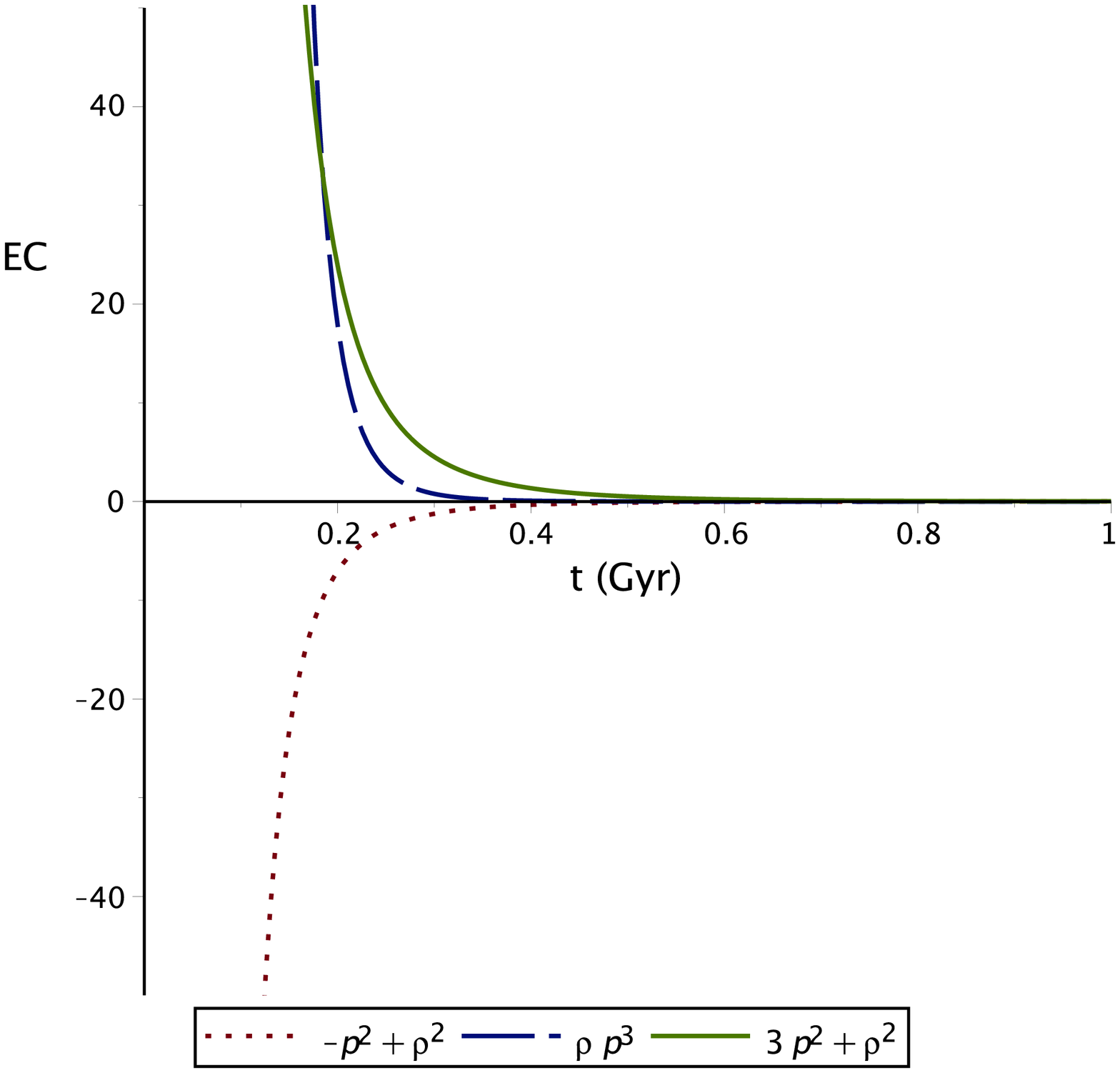}}
  \caption{The hybrid solution using $\Lambda(H)= \lambda +\alpha H + 3 \beta H^2 $: 
	(a) shows the sign flipping of $q$ from positive to negative. (b) The jerk parametr tends to $1$. (c) $p$ is always positive (d) The evolution of energy density. (e) and (f) show linear and nonlinear energy conditions. The sound speed causality condition $0 \leq C^2_s=\frac{dp}{d\rho}\leq 1$ is not satisfied for this model. Here $\alpha=7.2\times 10^{-17}$, $\lambda=\alpha=\beta=0.1$, $\alpha_1=\beta_1=0.$3}
  \label{fig:100}
\end{figure}

\section{Cosmographic Analysis}
The cosmography of the universe has recently become a popular research area \cite{cosmography1,cosmography2} in which cosmological parameters can be expressed in terms of kinematics only. As a result, cosmographic analysis is model-independent with no need to an equation of state to investigate cosmic dynamics \cite{cosmography3}. The Taylor expansion of the scale factor $a(t)$ around the present time $t_0$ is written as
\begin{equation} \label{taylor}
a(t)=a_0 \left[ 1+ \sum_{n=1}^{\infty} \frac{1}{n!} \frac{d^na}{dt^n} (t-t_0)^n \right]
\end{equation}
The following coefficients of the expansion (\ref{taylor}) are described respectively as the Hubble parameter $H$, deceleration parameter $q$, the jerk $j$, snap $s$, lerk $l$ and max-out $m$ parameters 
\begin{eqnarray} 
H=\frac{1}{a}\frac{da}{dt}~~,~~q=-\frac{1}{aH^2}\frac{d^2a}{dt^2}~~,~~j=\frac{1}{aH^3}\frac{d^3a}{dt^3}\\   \nonumber
s=\frac{1}{aH^4}\frac{d^4a}{dt^4}~~,~~l=\frac{1}{aH^5}\frac{d^5a}{dt^5}~~,~~m=\frac{1}{aH^6}\frac{d^6a}{dt^6}.
\end{eqnarray} 
For the first model we get
\begin{eqnarray} 
j&=& 1+\frac{2n^2-3n}{\cosh^2(\eta t)},\\
s&=&\frac{1}{\cosh^4(\eta t)} \left[\cosh^4(\eta t)-\cosh^2(\eta t)(4n^3-8n^2+6n)-2n^3+3n^2\right] \\
l&=&\frac{1}{\cosh^4(\eta t)} \left[\cosh^4(\eta t)-\cosh^2(\eta t)(8n^4-20n^3+20n^2-10n)+16n^4-30n^3+15n^2\right] \\
m&=&\frac{1}{\cosh^6(\eta t)} \left[\cosh^6(\eta t) +\cosh^4(\eta t) (-16n^5+48n^4-60n^3-15n) \right. \\  \nonumber
&+& \left. \cosh^2(\eta t)(-88n^5+196n^4-150n^3+45n^2)-16n^5+30n^4-15n^3 \right] \\
\end{eqnarray}
And for the second model
\begin{eqnarray} 
j&=& {\frac {{\alpha_1}^{3}+ \left( 3\,\beta\,t-3 \right) {\alpha_1}^{2}+
 \left( 3\,{\beta}^{2}{t}^{2}-3\,\beta\,t+2 \right) \alpha_1+{\beta}^{3}
{t}^{3}}{ \left( \beta\,t+\alpha_1 \right) ^{3}}},\\
s&=& \frac{1}{\left( \beta\,t+\alpha_1 \right) ^{4}} \left(\beta^4 t^4+4\alpha_1 \beta^3t^3+6 \beta^2 \alpha_1(\alpha_1-1) t^2+ (4\alpha_1^2-12\alpha_1+8)\alpha_1 \beta t \right. \\  \nonumber
&+& \left. \alpha_1^4-6\alpha_1^3+11\alpha_1^2-6\alpha_1  \right) \\
l&=& \frac{1}{\left( \beta\,t+\alpha_1 \right) ^{5}} \left( \beta^5 t^5 +5\alpha_1 \beta^4t^4 +10 \beta^3 \alpha_1 (\alpha_1-1) t^3 +10 \alpha_1 \beta_1 (\beta \alpha_1^2-3\beta \alpha_1+2\beta )t^2 \right. \\  \nonumber
&+& \left. 5\beta \alpha_1(\beta^3-6 \alpha_1^2+11 \alpha_1 -6)t+ \alpha_1^5 -10 \alpha_1^4+35 \alpha_1^3-50 \alpha_1^2 +24 \alpha_1\right)   \\
m&=& \frac{1}{\left( \beta\,t+\alpha_1 \right) ^{6}} \left( \beta^6 t^6 +6\alpha_1 \beta^5t^5 + 15 \beta^4 \alpha_1 (\alpha_1-1) t^4 +20 \alpha_1^2 \beta^3 (2\alpha_1^2+\alpha_1-3)t^3 \right. \\  \nonumber
&+& \left. 15\beta^2 \alpha_1(\alpha_1^3-6\alpha_1^2+11\alpha_1-6)t^2+6 \beta \alpha_1(\alpha_1^4-10\alpha_1^3+35\alpha_1^2-50\alpha_1+24)t \right. \\  \nonumber
&+& \left. \alpha_1^6 -15\alpha_1^5 +85\alpha_1^4-225\alpha_1^3+274\alpha_1^2-120\alpha_1 \right)   \\       
\end{eqnarray}
The value of $s$ is required for the study of the dark energy evolution. In addition to providing a suitable way to describe models close to $\Lambda CDM$, The presence of a transition time when cosmic expansion is modified is shown by the positive sign of $j$. The sign of $q$ also shows if the expansion is accelerating or decelerating. Despite its advantages, a valuable discussion of the cosmographic approach's limitations and downsides has been provided in \cite{cosmography1}. 
\section{positive $\Lambda$ and the eternal acceleration problem}
The behavior of the deceleration parameter $q$ in both models shows that it tends to a constant value $-1$ as $t \rightarrow \infty$. The same behavior happens for all cosmographic parameters $s$, $l$ and $m$ where they tend to a constant value at late-time. While $\Lambda$ represents the vacuum energy density and is supposed to has a tiny positive magnitude $\approx 10^{-123}$ \cite{1,1a}, It has been suggested that a negative $\Lambda$ can also fit a large data set and provide a solution to such 'eternal acceleration' problem which is a consequence of assuming a positive $\Lambda$ to explain the cosmic acceleration\cite{vinc}. The cosmological constant in the current model has a very small positive value at late-times and there's nothing to stop the accelerated expansion.

\begin{figure}[H]
  \centering            
	\subfigure[$s$]{\label{510}\includegraphics[width=0.3\textwidth]{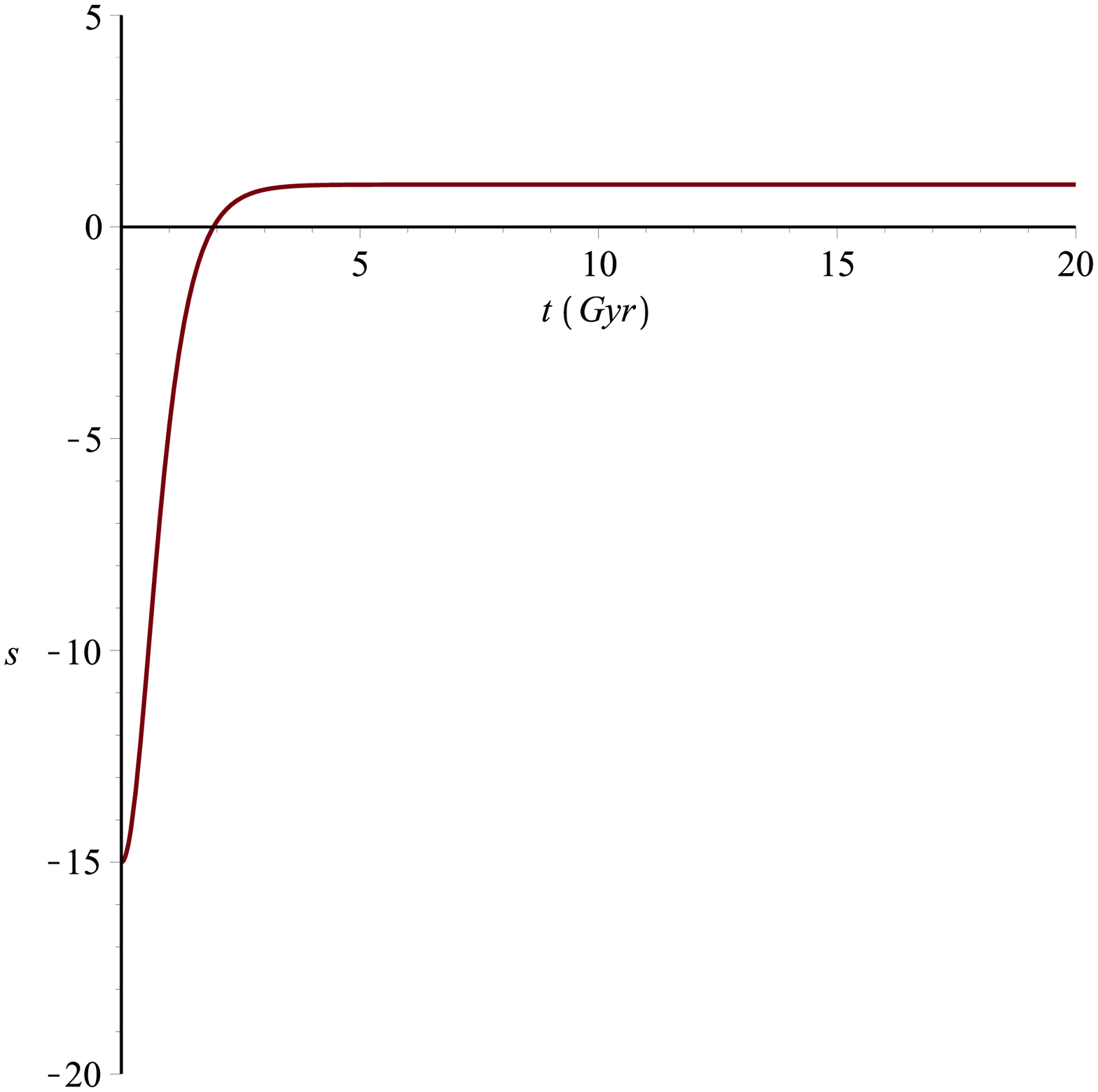}}
	\subfigure[$l$]{\label{610}\includegraphics[width=0.3\textwidth]{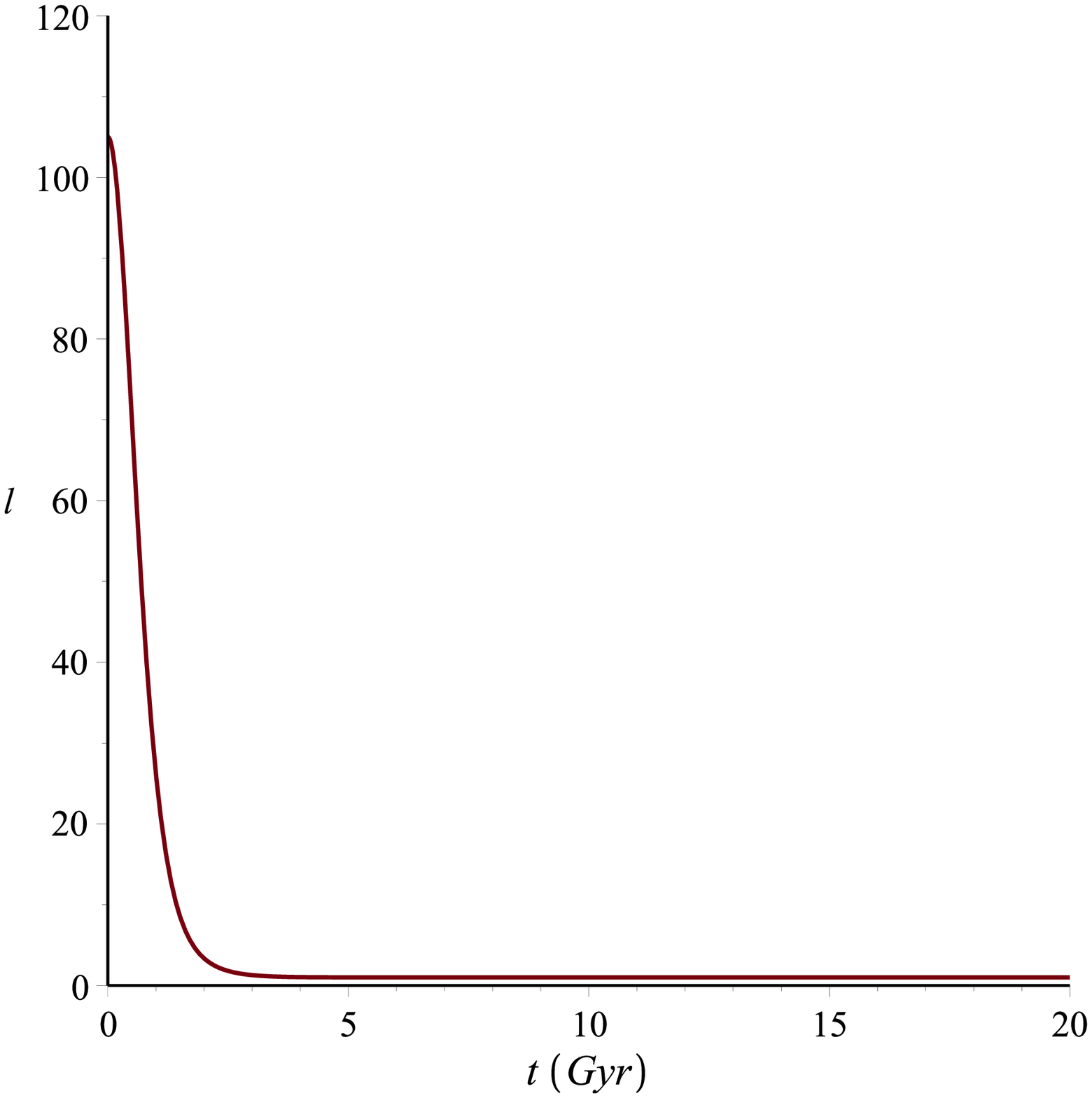}}
	\subfigure[$m$]{\label{710}\includegraphics[width=0.3\textwidth]{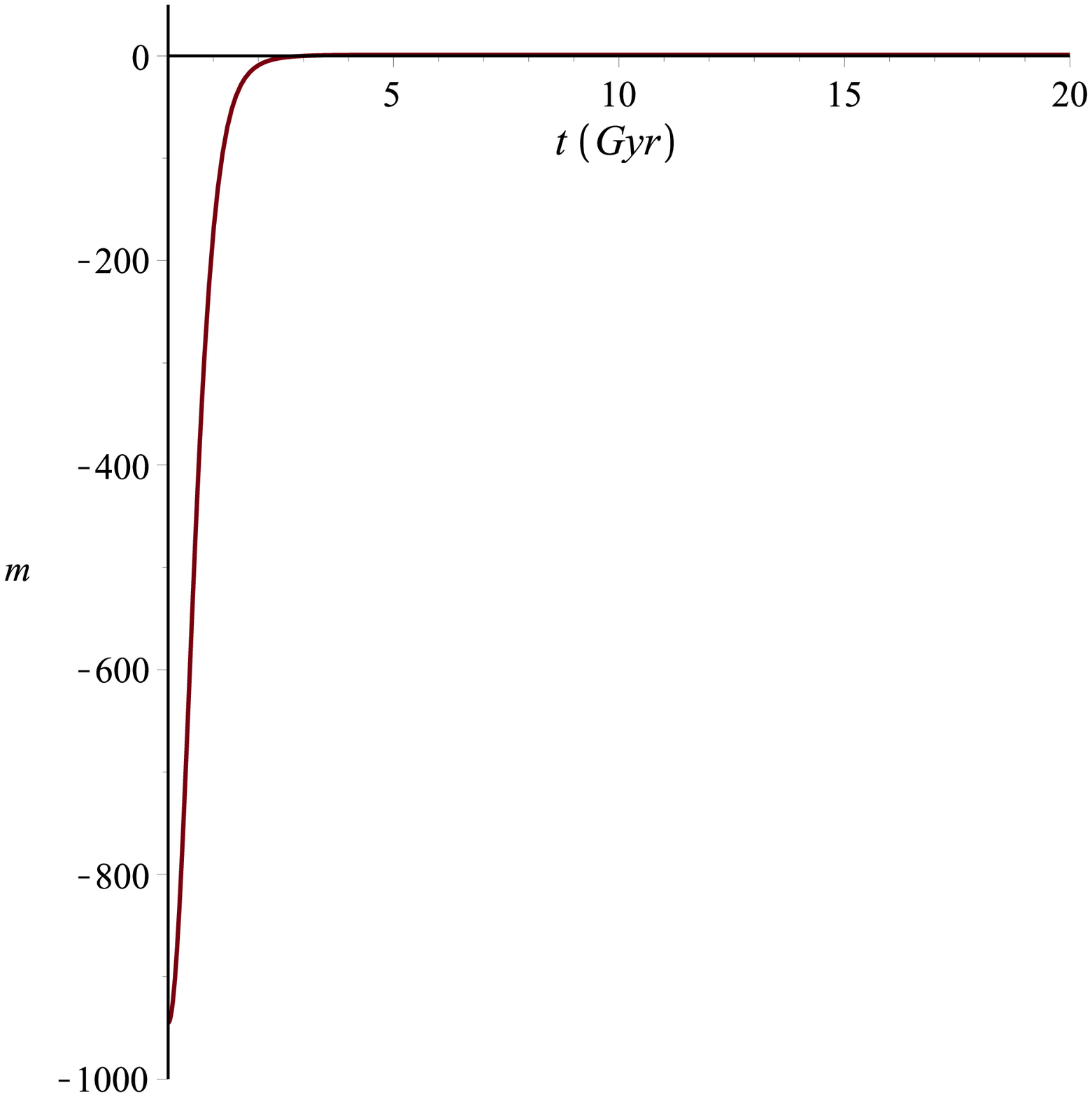}}\\
	\subfigure[$s$]{\label{810}\includegraphics[width=0.3\textwidth]{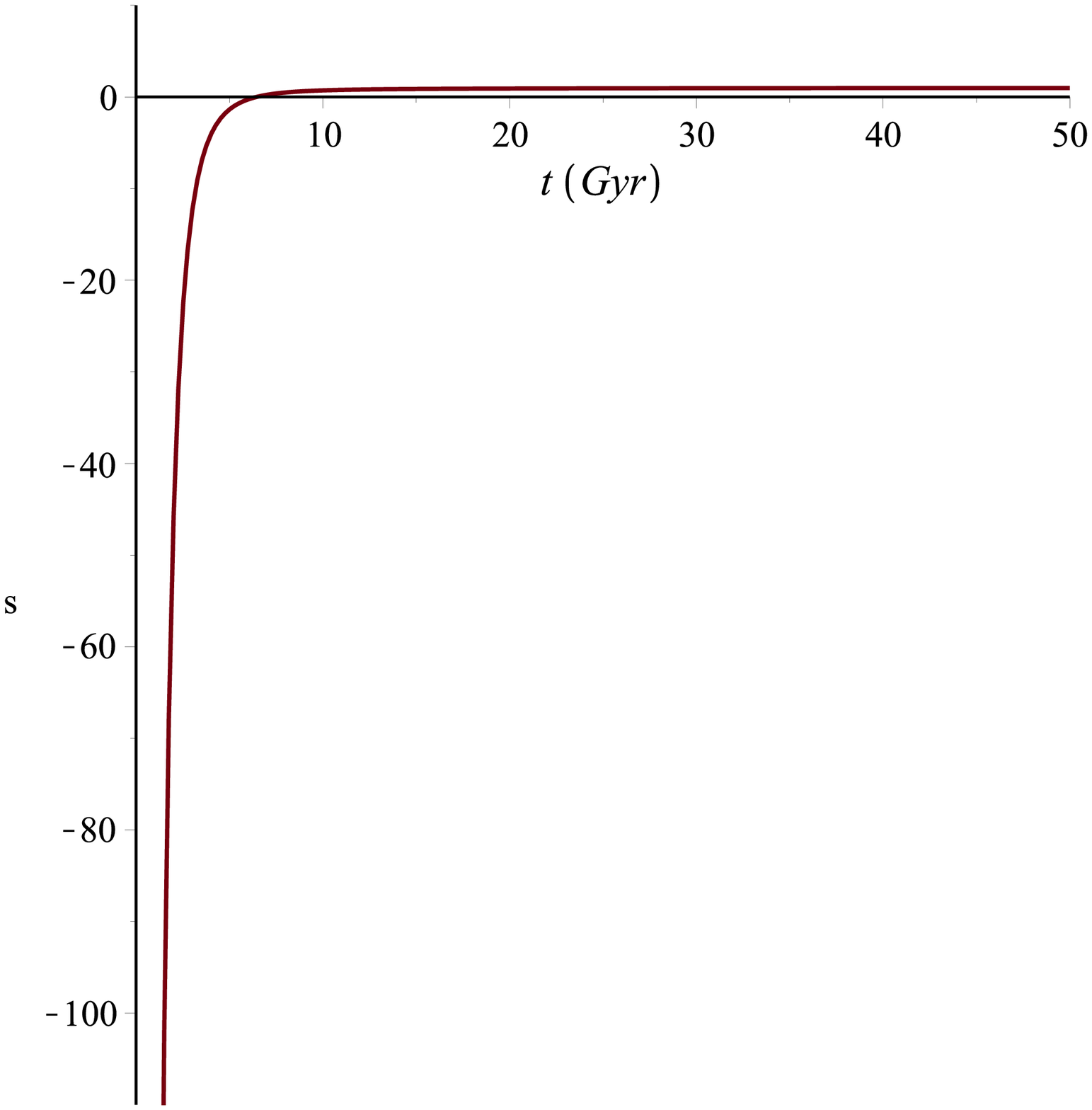}}
	\subfigure[$l$]{\label{910}\includegraphics[width=0.3\textwidth]{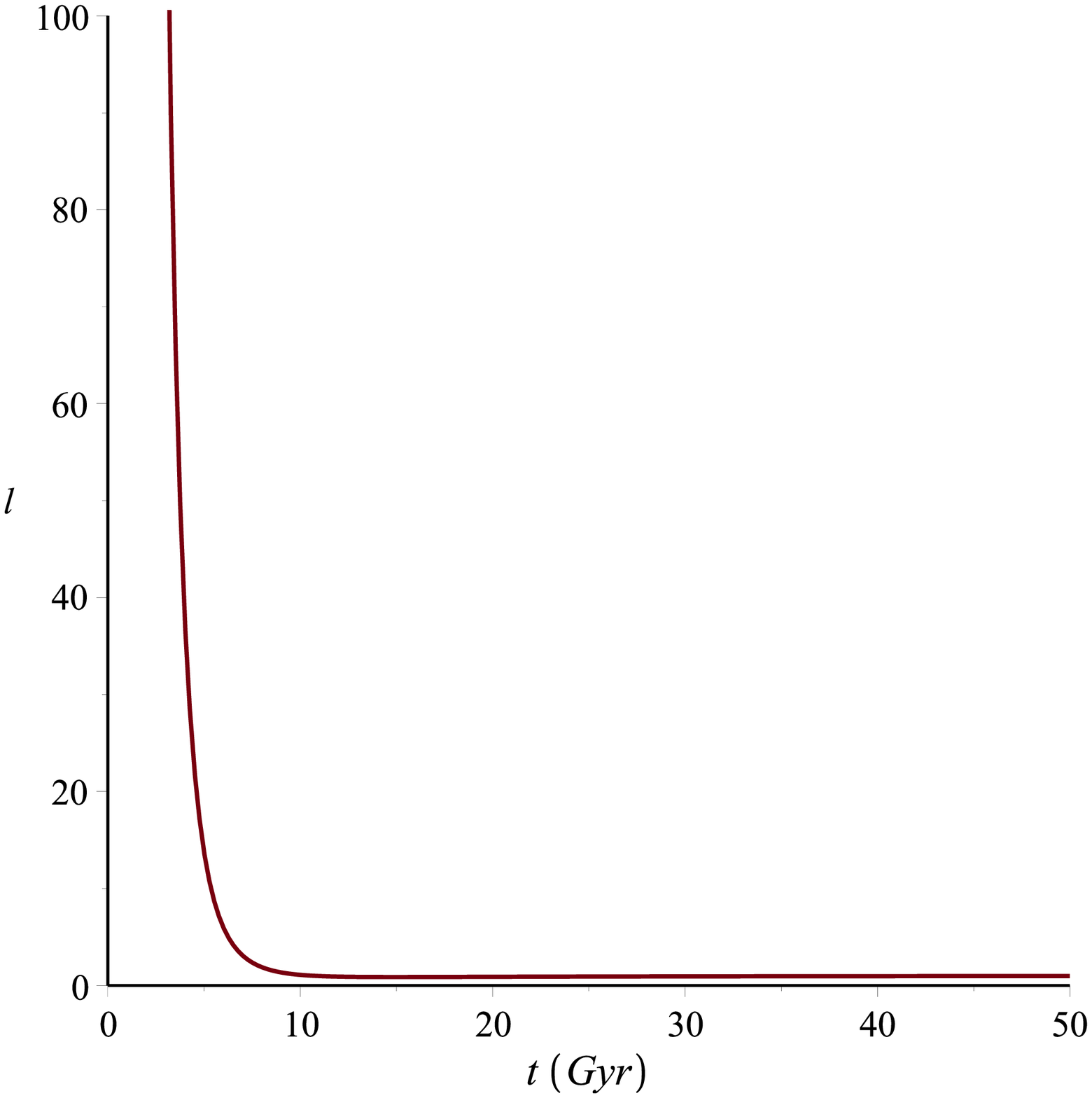}}
	\subfigure[$m$]{\label{1100}\includegraphics[width=0.3\textwidth]{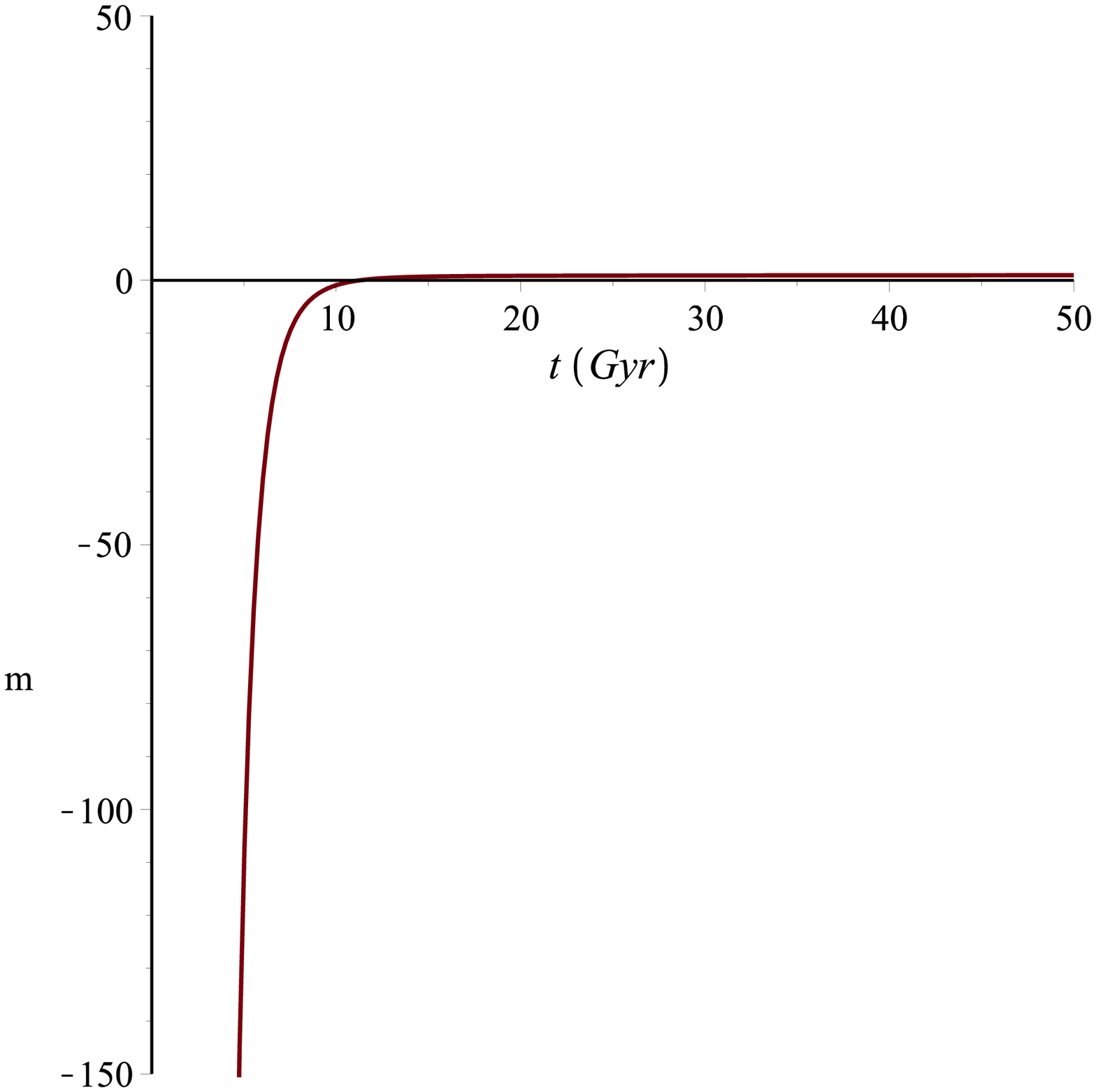}}
  \caption{Evolution of the cosmographic parameters $s$, $l$ and $m$ with cosmic time for model 1: (a), (b), (c) and for model 2: (d), (e), (f)  }
  \label{200}
\end{figure} 

\section{Conclusion}

Some cosmological features of the recently suggested 4D Gauss-Bonnet gravity have been investigated through two different models assuming a varying cosmological constant.  Unlike the usual viewpoint, In each model we have used a given scale factor derived from the desired cosmic behavior. The first hyperbolic model has been investigated through two different models of varying $\Lambda$: The first is $\Lambda (t) = \frac{C}{t^2}$ and the second is  $\Lambda(H)= \lambda +\alpha H + 3 \beta H^2 $. The second one gives better results where, for example, the sound speed causality condition along with the three nonlinear energy conditions are satisfied.  The deceleration-acceleration cosmic transit occurs in with a sign flipping in cosmic pressure from positive to negative. The deceleration parameter $q$ in both models tends to a constant value $-1$ as $t \rightarrow \infty$. Also, the jerk parameter also tends to $+1$ as $t \rightarrow \infty$ in both models.
The second hybrid model has been investigated through $\Lambda(H)= \lambda +\alpha H + 3 \beta H^2 $ where we have got a positive pressure with no sign flipping. Also, the sound speed causality condition is not satisfied.  

\section*{Acknowledgment}
We are so grateful to the reviewer for his valuable suggestions and comments that significantly improved the paper.

\end{document}